# Study of Black Phosphorus Using Angle-Resolved Polarized Raman Spectroscopy with 442 nm Excitation

**Weijun Luo[1], Qian Song[1], Guangnan Zhou[1], David Tuschel[2], and Guangrui Xia[1,*]**

[1]the University of British Columbia, Department of Materials Engineering, Vancouver, B.C., V6T1Z4, Canada
[2]HORIBA Scientific, HORIBA Scientific Inc, Edison, NJ, 08820, USA
[*]gxia@mail.ubc.ca

## ABSTRACT

We investigated 10 to 200 nm thick black phosphorus flakes on $SiO_2$/Si and polyimide substrates by Angle-resolved Polarized Raman spectra (ARPRS) using 442 nm excitation wavelength. The results revealed that ARPRS with 442 nm excitation can provide unambiguous, convenient, non-destructive and fast determination of BP's crystallographic orientation. The substrate and thickness dependencies of Raman spectra and Raman tensor elements were studied. These dependencies were shown to be influenced by Raman excitation laser heating effect. By comparing with in-situ Raman measurements at elevated temperatures, we were able to quantify the laser-heating effect, which is significant and hard to avoid for Raman measurements of BP on polyimide substrates due to the poor thermal conductivity of the substrate. Thermal processing by substrate heating was shown to have a significant impact on BP on $SiO_2$/Si substrate, but not for BP on polyimide due to smaller thermal expansion mismatch. Our results give important insights on Raman Spectroscopy characterizations of BP on different substrates.

## 1. Introduction

Semiconducting orthorhombic black phosphorus (BP) is a layer-stacked 2-dimensional (2D) nanomaterial with sp3-hybridized phosphorus atoms covalently bonded in-plane to a puckered honeycomb structure and the bonding between planes is via weak van der Waals forces[1] . Since the successful isolation of 2D BP reported in 2014,[2] many BP-based devices such as photo-detectors,[3] field-effect-transistors[4,5,6] , solar cells have been successfully demonstrated, which are promising especially for applications in flexible and ultra-thin electronics and photonics[2,7,8,9,10,11] . BP's physical properties such as mechanical property[12,13] , carrier mobility[14,15] , optical and phonon properties[16,17,18,19,20] all exhibit strong crystal orientation dependence. Moreover, the energy band structure of BP (with a direct band gap at the Γ point of the Brillouin zone[7]) strongly depends on its thickness. The band gap varies from 2 eV for a monolayer[21] to 0.3 eV for bulk BP ( > 20 atomic layers,[22] or about 10 nm[15]). Hence, developing a convenient, accurate and non-destructive method to quickly determine BP's crystal orientation and thickness is critical for BP's device applications.

On the crystal orientation determination, angle-resolved conductance of BP sheet[14,18,23] , diffraction pattern by high-resolution transmission electron microscope[17,24] , optical absorption[14,17] and Angle-resolved Polarized Raman Spectroscopy (ARPRS) have been used to determine the in-plane crystal orientation of BP[16−19] . Moreover, ARPRS is also a good way to study the anisotropic electron-phonon interaction in BP, which is important for electrical, thermal and super-conductive properties. However, the ARPRS response of BP is quite complicated, which depends on the Raman excitation laser wavelength[16,17,19] . For instance, using a 532 nm excitation laser, Ribeiro[19] et al reported their observations of a relative larger maximum peak intensity of $A^2_g$ mode when the excitation laser polarization was parallel to the zigzag direction in relation to a relative smaller maximum peak intensity when excitation polarization was along the armchair direction, which was opposite to the observation using 514.5 nm excitation wavelength by Wu[18] et al. What's more, even with the same excitation wavelengths (532, 633 and 785 nm), observations by Ling[17] et al using samples of different thickness showed that the main axis of Raman intensity polar diagrams measured by ARPRS can be either along the zigzag or the armchair direction depending on the BP thickness and the excitation wavelength. Their work systematically analyzed the complications in BP ARPRS results using quantum perturbation theory, semi-classical model of polarized Raman scattering and group theory. The ratio between the amplitudes of two Raman tensor element c and a, $|\frac{c}{a}|$, varies from less than 1 to more than 1 with the excitation wavelength and the thickness, which is responsible for the change of the main axis direction. They also explained the phase difference between a and c ($\varphi_{ac}$) determines the positions of the minimum peak intensity of $A^1_g$ and $A^2_g$ modes, and affects the shape of the ARPRS polar diagrams. However, their study didn't used shorter excitation wavelengths, which turned out to be more convenient in BP orientation determination.

Ribeiro et al[25] reported ARPRS results of a 360-nm-thick BP sample using 532 and 488 nm excitations. They first calculated the magnitudes of a, c, and $\varphi_{ac}$ for $A^1_g$ and $A^2_g$ mode. The results show that $\varphi_{ac}$ for $A^1_g$ is zero for these two wavelengths, but $\varphi_{ac}$ for $A^2_g$ is not zero. A study by Kim et al.[16] investigated the Raman excitation wavelength and BP thickness dependence of the ARPRS response using 442, 488, 514, 532 and 633 nm excitations, and the BP thickness ranged from 5 nm to 90 nm. They demonstrated that among those wavelengths, ARPRS with 442 nm excitation is the most convenient way to determine the BP crystal orientation. The reason lies in the fact that $\varphi_{ac}$ of $A_g^1$ mode is zero for 442 nm excitation, and non-zero for other wavelengths they studied. This fact means that there exist only two maxima of the BP $A_g^1$ Raman intensity in a polar diagram, where the intensity is plotted as a function of the rotation angle $\theta$ from 0 to 360° ($\theta = 0°$ is defined at when excitation polarization is parallel to the zigzag direction). The two maxima were at $\theta = 0°$ and $\theta = 180°$ and the two minima exist at $\theta = 90°$ and $\theta = 270°$, forming a bow-tie shape in Raman polar plots of $A_g^1$ mode using a 442 nm excitation. $A_g^2$ mode polar diagrams have a similar bow-tie shape, but is perpendicular to that of the $A_g^1$ mode bow-tie. With 442 nm excitation, the relative intensity ratio between the maxima and the minima Raman intensity is $\frac{|c_1|^2}{|a_1|^2}$ for $A_g^1$ mode and $\frac{|a_2|^2}{|c_2|^2}$ for $A_g^2$ mode independent of BP thickness. On the contrary, for all other wavelengths they studied, the relative intensity ratio between the maxima and the minima depends on the BP thickness, which can be one for certain thickness and wavelength, making it hard to determine the main axis of the BP polar diagrams and thus the crystal orientations. The key literature results we quoted above were those with parallel polarization configuration, and on SiO2/Si substrates[16,17,25].

As discussed above, Raman tensor elements a and c as well as their phase difference $\varphi_{ac}$ for $A_g^1$ and $A_g^2$ determine the shape and relative intensity in ARPRS polar plots, which are important parameters to obtain. In the semi-classical model of polarized Raman scattering theory, they are the partial derivatives of the dielectric constant of BP with respect to the normal coordinates of the corresponding phonon modes along xx and zz directions respectively. By convention, xx and zz refer to the armchair and zigzag direction respectively. As the dielectric constant of BP depends on the BP thickness, a, c and $\varphi_{ac}$ should also depend on the BP thickness besides the excitation wavelength dependence. a, c and $\varphi_{ac}$ vs. thickness were only available for 633 nm excitation in Ling et al's work.[17] For 442 nm excitation, a, c and $\varphi_{ac}$ values were available for 90 nm BP only.[16] The literature values were summarized in Table S13.

Up to date, few-layer BP preparation relies on thinning from bulk BP samples. Thermal sublimation thinning has been realized, and BP flakes can be in-situ monitored by Raman spectroscopy during sublimation.[26] Therefore, it is very important to establish a relation between BP Raman response and its thickness for fast determination of BP thickness instead of ex-situ thickness measurements. In this work, using excitation wavelength of 442 nm under parallel configuration, we investigated the ARPRS response of 27 BP samples on 300 nm SiO2/Si substrates and on 100 $\mu$m polyimide substrates prepared by sublimation thinning and mechanical exfoliation. The BP thickness ranged from 15 nm / 10 nm to 195 nm / 200 nm for these two types of substrates respectively. It was confirmed that, with 442 nm excitation, ARPRS provides a fast and unambiguous way to determine BP crystallographic orientation. Based on the ARPRS polar diagrams measured, we calculated the magnitudes of BP's Raman tensor elements a and c and $\varphi_{ac}$ as a function of BP thickness after eliminating the anisotropic interference effect. Furthermore, we performed in-situ Raman measurements at elevated temperatures at two thick BP flakes (thickness > 200 nm) on the two types of substrates. By comparison, we quantified that the laser-heating effect from Raman excitation laser, which is significant unavoidable for BP on polyimide substrates. For BP on SiO2/Si, the laser heating effect is insignificant. However, thermal processing by substrate heating can cause a strain change during heating, which is a practical concern.

## 2. Results and discussion

In the first experiment, BP flakes of various thickness were measured by ARPRS with 442 nm excitation at room temperature. The samples were either on 300 nm SiO2/Si substrates or on flexible polyimide substrates. The BP samples were obtained from bulk BP by mechanical exfoliation and/or thermal thinning.[26] The thickness were measured by atomic force microscopy (AFM). The ARPRS measurements were in the backscattering geometry, under parallel configuration. BP samples were placed on a rotation stage mounted (on the motorized stage of Raman microscope) that can rotate 360 degrees in-plane. In the second experiment, thick BP flakes (> 200 nm) on these two types of substrates were placed on a heating stage and Raman spectra were collected at elevated temperatures from 25 to 200 °C, the results of which were used to provide temperature dependent Raman spectra to help explaining the results in the first experiment. The second experiment is also referred to as the thermal heating experiment below.

### 2.1 Unambiguous determination of BP's crystallographic orientation using 442 nm excitation ARPRS

Let's look at the room temperature Raman spectra first. In Figure 1, we listed the results of a 25-nm-thick BP sample on SiO2/Si and a 10-nm-thick BP on polyimide. The crystal orientation identification (zigzag and armchair directions) followed previous work[16,18]: in a polar diagram, the direction along which the maximum Raman peak intensity of $A_g^1$ is obtained represents the zigzag direction while that of $A_g^2$ represents the armchair direction. $B_{2g}$ mode is much weaker than $A_g^1$ and $A_g^2$ modes[17,18,25]. BP spectra in Figure 1.(B) show typical lattice vibrational modes $A_g^1$ and $A_g^2$ at 362.17 cm$^{-1}$ and 466.97 cm$^{-1}$, which agrees with previous work[16,24]. The peak at 520.7 cm$^{-1}$ is the Si peak; in Figure 1.(F), 10-nm-thick BP sample on polyimide shows around 361.36 cm$^{-1}$ and 465.64 cm$^{-1}$, and in comparison with 25-nm-thick BP flake on SiO$_2$/Si, there is about one wavenumber shift in the Raman peaks. This observation is similarly to BP thin films on polyethylene terephthalate (PET), which has Raman peak positions two wavenumbers less than those on SiO$_2$/Si substrates.[8] The discrepancy of peak positions of BP on different substrates can be attributed to the laser heating effect: the heat dissipation is much slower for BP on polyimide substrate than BP on SiO$_2$/Si substrate. In Section 1 of Supporting information, we demonstrated that, in our study, when the laser power was reduced to 0.2 mW, the temperature of BP sample on polyimide substrate was increased while that of BP sample on SiO$_2$/Si substrate was not; moreover, we illustrated that the Raman spectra of bulk BP sample on polyimide substrate at room temperature matched with that of bulk BP sample on SiO$_2$/Si substrate measured at 60 °C, while Raman spectrum of 10-nm-thick BP on polyimide substrate agrees with bulk BP (>200 nm) measured at 100 °C (See Figure 5.(A) and (G)). This heating effect for BP on polyimide is hard to avoid as more light intensity attenuation would result in Raman signal intensity too weak to be measured. The laser heating effect is discussed in more detail later in Section 2.3.1.

In Figure 1.(C), (G), (K) and (N), Raman intensity polar diagrams of A$^1_g$ and A$^2_g$ modes for BP samples on both types of substrates are listed. The angle values of the polar diagrams are the angle readings of the rotation stage. Therefore, these polar diagrams were not aligned to any specific crystal orientation. As A$^1_g$ and A$^2_g$ were measured simultaneously, two data points on A$^1_g$ and A$^2_g$ polar diagrams of one sample with the same angle readings mean that these two data points are from the same spectrum. The main axes of the polar diagrams are shown by the double-headed arrows. It can be seen that for all the thicknesses studied (10 to 200 nm), all the polar diagrams have the bow-tie shape with two maxima and the main axes of A$^1_g$ and A$^2_g$ modes of the same sample are always perpendicular to each other. Moreover, there are no significant changes in the ratios between the maxima and the minima, $\frac{|c|^2}{|a|^2}$ which can make a bow-tie shape to be more like a peanut shape. Compared to other excitation wavelengths such as 488, 514.5, 532, 633 and 785 nm[16,17], the Raman polar diagrams with 442 nm excitation are much more cleaner with no secondary maxima or a significant change in $\frac{|c|^2}{|a|^2}$ or changes in the angle between the A$^1_g$ main axis and A$^2_g$ main axis.

Therefore, with expanded thickness ranges and the addition of polyimide substrates, it is confirmed that ARPRS with 442 nm excitation can provide unambiguous and convenient determination of BP's crystal orientations. More ARPRS data and polar diagrams of different BP samples are listed in Table S7 and S8 of Section 3 of Supporting Information.

Here we suggest the much more cleaner Raman polar diagrams with 442 nm excitation can be attributed the fact that the resonant Raman scattering[27] took place when excitation polarization was parallel to BP's armchair direction: the laser energy (2.8 eV for 442 nm) is resonant to an electron transition from a valence band state to a conduction band state[19,28] resulting a very strong Raman intensity. This can be seen in the much stronger A$^2_g$ peak in Raman spectra when excitation polarization is parallel to armchair direction in Figure 1.(F) and (M), as well as Table S7 and S8 in Supporting information. Considering the given band structure[17] of bulk BP and character tables[17,25] (listed in Table 1), we suggested the transition from B$_{3g}$ → B$_{2u}$ is in resonance with 442 nm excitation laser. Theoretical studies calculated that the energy difference between B$_{3g}$ and B$_{2u}$ states are 2.5 and 2.8 eV respectively[17,29]. Our measurements support the later value.

## 2.2 Raman tensor elements a and c, and $\varphi_{ac}$

After the ARPRS measurements, the next step was to calculate the Raman tensor elements a and c as a function of BP thickness and substrate type. First, we need to rule out the interference effect[16,17] due to multiple reflections in BP/substrate layers. (See details of calculation in Section 5 in Supporting Information)

As discussed previously, $|\frac{c}{a}|^2$ is from the ratio between the Raman peak intensity measured along the zigzag direction and that measured along the armchair direction in a polar diagram. The values of $|\frac{c}{a}|$ for A$^1_g$ and A$^2_g$ are plotted in Figure 2. As Raman peak intensity was very low when excitation polarization was parallel to armchair direction of BP, $|\frac{c_1}{a_1}|$ of A$^1_g$ mode was more noisy. Notably, regardless of thickness and substrates, $|\frac{c_1}{a_1}|$ of A$^1_g$ mode here is always larger than 1 which means the maximum peak intensity of A$^1_g$ appears when excitation polarization is parallel to the zigzag direction. For A$^2_g$ mode, $|\frac{c_2}{a_2}|$ of A$^1_g$ is always less than 1, that, A$^2_g$ peak reaches its maximum intensity when the excitation polarization is parallel to the armchair direction. An interesting observation is that for BP thickness larger than 10 nm, we observed that $|\frac{c_2}{a_2}|$ of A$^2_g$ mode monotonically decreased with increasing thickness, and the thickness dependence of A$^1_g$ was weaker. The thickness dependence will be discussed in Section 2.3.

In comparison, the values of $|\frac{c}{a}|$ fluctuate above or below 1 when using longer excitation wavelength like 532 nm, 633 nm and 785 nm, leading to the uncertainty determining the main axis[17]. Therefore, using excitation wavelength of 442 nm provides definite identification of BP's crystallographic orientation. Meanwhile, for all the sample measured with the $|\frac{c}{a}|$ varies from 1.2 (0.09) to 3.7 (0.3) for the A$^1_g$ (A$^2_g$) mode, no secondary intensity maxima were observed. This means that the phase difference $\Phi_{ca}$ in our results is zero, independent of the sample thickness and substrate, which is in agreement with previous work.[16]

In addition, we list the calculated amplitudes of Raman tensors a and c of both $A_g^1$ and $A_g^2$ modes as functions of thickness of BP in Table S14 of Supporting Information. Despite the substrate types, amplitudes of a and c of $A_g^1$ and $A_g^2$ modes all decrease with increasing thickness.

## 2.3 Substrate and BP thickness dependence

The energy band gap ($E_g$) has been studied well by experiments, and it was shown that $E_g$ stabilizes as the BP thickness goes above 20 layers ($\approx$ 10 nm)[22] . Raman response of a material involves photo-electron and electron-phonon interactions, which are determined by the energy band structure. Therefore, Raman provides an indirect way of studying the energy band structure.

In this study, after the determination of crystal orientation, we collected Raman spectra for BP on two different substrates with a thickness (t) range from 10 to 200 nm, which were all measured with their zigzag direction parallel to the excitation polarization. Raman peak positions, the full width half maximum (FWHM) and peak intensity ratio of $\frac{A_g^1}{A_g^2}$ were plotted as a function of the thickness (Figure 3.). The intensity ratio of $\frac{A_g^1}{A_g^2}$ for this direction is also referred as max($\frac{A_g^1}{A_g^2}$) in this paper as $\frac{A_g^1}{A_g^2}$ ratio reaches maximum along the zigzag direction. From Figure 3. (A), we can see that the peak positions of BP on $SiO_2$/Si is stable with the thickness, but decrease with decreasing thickness for BP on polyimide, showing a larger heating effect for BP on polyimide especially for thinner BP. In Figure 3. (B) and (C), FWHM of both peaks decrease with increasing thickness, which stabilizes around 50 nm. Figure 3.(D) shows the Raman intensity ratio $\frac{A_g^1}{A_g^2}$ as a function thickness, which is $\frac{|c_1|^2}{|c_2|^2}$ after removing the interference effect. The ratios for both substrate types increase with increasing thickness and stabilize above 100 nm. The thickness dependence of $\frac{A_g^1}{A_g^2}$ for BP on $SiO_2$/Si was stronger than that for BP on polyimide. These thickness dependencies were against our initial expectations for t > 10 nm range, which were commonly considered as bulk BP thickness range. For t > 10 nm, we expected that 1) Raman results should be independent of the substrate type and 2) there should not be a thickness dependence. To explain the unexpected results, several factors were considered: 1) temperature difference due to different thermal dissipation capability of the sample and the substrate under Raman excitation laser heating, which has been investigated widely in graphene,[30] $MoS_2$[31] and carbon nanotubes;[32] 2) thinner BP samples were primarily prepared by thermal sublimation, during which strain can be introduced; 3) crystal quality change due to the air instability and light-enhanced degradation of BP;[33] and 4) energy band structure dependence on thickness. Few studies are available on the energy band structure other than the energy bandgap measurements. Therefore, in the following discussion, we will focus on the first three factors.

### 2.3.1 Temperature effect by laser heating

To establish the temperature dependence, in the second experiment, we put a thick BP flake (> 200nm, on a $SiO_2$/Si substrate) on a heating stage and measured Raman spectra in-situ when it was heated from room temperature up to 200 °C. In Figure 4, Raman peak positions and $\frac{A_g^1}{A_g^2}$ were plotted as a function of the heating temperature. With increasing temperature, Raman peak positions decreased with a rate of -0.0156 $cm^{-1}$/K and -0.0358 $cm^{-1}$/K for $A_g^1$ and $A_g^2$ peak respectively. This trend was in accordance with the observations from previous works[34,35] ; $\frac{A_g^1}{A_g^2}$ decreased with increasing temperature. The spectra under heating were then used as a temperature probe to determine the BP temperature during the measurements. Similar temperature-dependent Raman spectra of another BP flake on polyimide were collected and shown in Figure 4.(A). What's more, with increasing temperature, Raman peak positions decreased with a rate of -0.0218 $cm^{-1}$/K and -0.0295 $cm^{-1}$/K for $A_g^1$ and $A_g^2$ peak, respectively, which was similar to temperature-dependent Raman spectra of BP on SiN membrane.[23]

In Figure 5. (A) to (G), we compared the Raman spectra of BP samples on polyimide substrates with various thickness with the bulk BP sample on $SiO_2$/Si measured at elevated temperatures. According to Table 2., for each thickness of BP on polyimide, we were able to find a temperature at which Raman peak positions, FWHM and $\frac{A_g^1}{A_g^2}$ of BP on $SiO_2$/Si could match very well. For example, the spectrum of 10-nm-thick BP on polyimide matches bulk BP on $SiO_2$/Si measured with 100 °C heating and that of 150-nm-thick BP on polyimide matches bulk BP on $SiO_2$/Si measured with 60 °C heating. The best fitting temperature goes up monotonically with decreasing BP thickness, showing that thinner BP on polyimide is more susceptible to the laser heating effect. This is consistent with the Raman peak red-shifts with decreasing thickness for BP on polyimide. As Raman peak positions, FWHM and $\frac{A_g^1}{A_g^2}$ are sensitive to the four factors we discussed above, being able to match BP on polyimide with bulk BP on $SiO_2$/Si suggests that they have the close temperature, strain, crystal quality and energy band structure. Therefore, we can attribute the Raman thickness dependence of BP on polyimide to the laser heating effect which has a BP thickness dependence. For a certain laser power input and substrate,the thickness dependence of $\frac{|c_2|}{|a_2|}$ and $\frac{|c_1|^2}{|c_2|^2}$ can be established such as in Figure 2.(B) and Figure 3.(D), which can be used for BP thickness determination.

The thickness dependence shown in Figure 3. for BP on $SiO_2/Si$ is more complicated. First, from Figure 3.(A), BP samples < 50 nm on $SiO_2/Si$ have BP $A_g$ mode Raman peaks blue-shifted. Thinner BP has more blue shift. A blue shift in Raman peaks can be a result of lower temperature[36] or compressive strain[37] . In Figure 8.(A), from Raman spectrum of a 15 nm BP on $SiO_2/Si$, the substrate Si peak is not changed from 520.7 $cm^{-1}$, which is the peak position from a blanket Si substrate during calibration even with the lowest excitation laser intensity. Therefore, we believe that BP temperature stayed at room temperature during the measurements.

### 2.3.2 Strain change during thermal processing

In Figure 6.(A), we compared the Raman spectra of those 2 BP samples on two different substrate types before and after the thermal heating experiment. Obviously, for BP on $SiO_2/Si$ substrate, $A_g^1$ and $A_g^2$ peak positions underwent a red shift after the thermal heating experiment; however, those of BP on polyimide came back to the pristine positions after cooling down to room temperature. In addition, the FWHM of $A_g^1$ and $A_g^2$ modes of BP on $SiO_2/Si$ exhibit a significant change, while those of BP on polyimide substrate retains. Considering all spectra were taken at room temperature and with same settings, we suspect that a strain change during thermal heating by film slipping resulted in the changes of peak shapes and positions of BP on $SiO_2/Si$ substrate. This cannot be attributed to thermal expansion mismatch strain as the temperature of the starting point and the ending point are the same. The strain change during heating such as film slipping can be induced by a thermal expansion mismatch strain at higher temperatures, and is suspected to be thickness dependent to explain the thickness dependence shown in Figure 5. For BP on polyimide, the thermal expansion mismatch between the BP films and the substrates are less than those on $SiO_2/Si$, which may explain the lack of strain change during heating. Please refer to Table S1 in the Supporting Information on the thermal expansion coefficients of BP, Si and polyimide. For BP on polyimide, the observation suggests that there are no strain change during heating.

### 2.3.3 Degradation of film quality during ARPRS measurements

Another possible reason may come from the photo-assisted oxidation and reaction with moisture[33] of BP during long-time exposure in air ambient during the measurements, which could degrade the crystal quality, then result in the changes of peak shapes and positions. We plotted the FWHM of $A_g^1$ and $A_g^2$ modes as functions of rotation angle instead of using integrated intensity in Fig.(A) and (B) of Figure 7, which was used for evaluating the degradation of crystal with increasing time as the measurements were performed from 0 to 360 degrees with 15 degrees as the interval. In agreement with Fig.(B) and (C) of Figure 3, the FWHM increases for thinner samples. However, by comparing the FWHM of $A_g^1(A_g^2)$ mode at the starting and ending point, no significant changes were observed. Hence, it could be concluded that in our work, the degradation might affect the peak shapes and positions of $A_g^1$ and $A_g^2$ modes, but it was not a significant factor.

Although Raman spectra can be influenced by these factors, this work is the first systematic study on the BP thickness dependence with 442 nm excitation. Future work needs to be performed to reduce or quantify these effects to give a clearer picture on the thickness dependence for the ultimate goal of quick orientation and thickness determination with Raman spectroscopy.

## 4. Conclusions

In summary, we investigated 10 to 200 nm thick black phosphorus flakes on $SiO_2/Si$ and polyimide substrates by ARPRS using 442 nm excitation wavelength. The results revealed that ARPRS with 442 nm excitation can provide unambiguous, convenient, non-destructive and fast determination of BP's crystallographic orientation. The substrate and thickness dependencies of Raman spectra and Raman tensor elements were studied. These dependencies were shown to be influenced by Raman excitation laser heating effect and substrate thermal conductivity. By comparing with in-situ Raman measurements at elevated temperatures, we were able to quantify the laser-heating effect, which is significant and hard to avoid for Raman measurements of BP on polyimide substrates due to the poor thermal conductivity of the substrate. Thermal processing by substrate heating was shown to have a significant impact on BP on $SiO_2/Si$ substrate, but not for BP on polyimide due to smaller thermal expansion mismatch. Our results give important insights on Raman Spectroscopy characterizations of BP on different substrates.

## 5. Experimental methods

*Sample preparation.* BP flakes were first mechanically exfoliated from bulk BP crystals (99.998 %, Smart Elements, Vienna, Austria) in a glove box and transferred to a pre-cleaned Si (100) wafer (4 inch, 0.56 mm thick) with a 300 nm thick thermal $SiO_2$ layer and a piece of 100 $\mu m$ thick polyimide film (DuPont$^{TM}$ Kapton HN general-purpose film, 100 $\mu m$ thick. The surface roughnesses of $SiO_2/Si$ and polyimide were around 0.3 nm and 1 nm, respectively. The wafer and polyimide were then cut into small pieces (3 mm×3 mm). In order to prepare thin BP flakes with various thickness less than 50 nm, we performed thermal sublimation (sample to be annealed at 230°C) on some exfoliated BP flakes to reduce their thickness. In our study, those BP samples on SiO$_2$/Si with thickness less than 45 nm were prepared by thermal sublimation; BP samples on polyimide substrate with thickness of: 10 - 40 nm, 58 - 70 nm and 85 nm are prepared by thermal sublimation method.

*Angle-resolved Polarized Raman spectroscopy (ARPRS).* Raman spectra were collected on a backscattered Horiba Jobin Yvon HR800 Raman system with 442 nm (2.81 eV) line from a He-Cd laser. The 442 nm incident laser beam was polarized, and a polarization analyzer was placed in parallel configuration (incidence and scattering are parallel) between the notch filter and the entrance of detector.[18] In this work, all ARPRS experiments were conducted under the parallel configuration following previous work[16-18,25]. Hereafter, considering that polyimide substrate was damaged by 442 laser without reducing its power (around 2 mW/$\mu m^2$), notch filter D=1 was then chosen in all ARPRS experiments in this work to lower the laser power by 10 times in order to avoid damage on both the samples and polyimide substrate. The acquisition time and accumulation time were optimized in order to enhance the Raman signals of BP samples as well as minimize the laser damage to BP samples. BP samples were rotated 360° about the microscope optical axis in 24 steps (15°/step). During every step, Raman measurement was performed at the same point on each sample to ensure the consistency of results. The grating number of detector was set as 2400 and spectral range from 300 to 600 cm$^{-1}$. The spectral resolution before fitting was ≈ 0.27 cm$^{-1}$.

*Thermal heating experiment* Typical procedure for collecting Raman spectra during thermal heating experiment can be described as: 1. by using ARPRS mentioned above, the excitation polarization was aligned at the Zigzag direction of thick BP flake (thickness > 200 nm) sample on SiO$_2$/Si (Polyimide) substrate (See details in Table S2 of supporting information); 2. Then the sample was carefully transferred to a Linkam TS1500 heating stage, which was mounted on the motorized XY stage of the Raman microscope; 3. with nitrogen (N$_2$) purging the heating stage, the sample was heated up from room temperature to 300 °C with a rate of 10 °C/min, and during every 10 °C, the temperature was held for 5 minutes in order to collect Raman spectra.

*Atomic Force Microscopy (AFM).* AFM measurements were performed using contact mode by an Asylum Research Molecular Force Probe 3D atomic force microscope and a Bruker Atomic Force Microscopy System.

## Author contributions

W. L. and G. X. initiated the project; W.L designed the experiments; W. L. prepared the samples; W. L. performed AFM measurements. W. L and Q. S. conducted the Raman measurements; W. L. and Q. S. analyzed the results; W.L. and G. X. led the writing of the paper. All authors discussed the results and contributed to the writing and revision of the manuscript. The whole project was supervised by G. X..

## Acknowledgment

W.L. acknowledges Professor. Hyeonsik Cheong (Department of Physics, Sogang University, Seoul, Korea.) for helpful discussions on the total enhancement factor of the interference effect. W.L. acknowledges Dr. Shengxi Huang (Department of Electrical Engineering and Computer Science, Massachusetts Institute of Technology.) for helpful discussions on anisotropic electron-phonon interaction. W.L. acknowledges Joshua.T.Cantin (Department of Chemistry, the University of British Columbia, Vancouver, Canada) for the discussion in internal conversion rate and different coupling of states. W.L. acknowledges Chris Balicki (4D LABS, Simon Fraser University, Burnaby, BC, Canada) for his assistance in the use of AFM.

## Additional information

**Supplementary information:**

**Competing financial interests:** All other authors declare no competing financial interests

| Optical image | Raman spectrum | Polar diagram of $A_g^1$ mode |
|---|---|---|

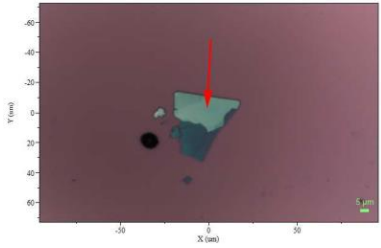

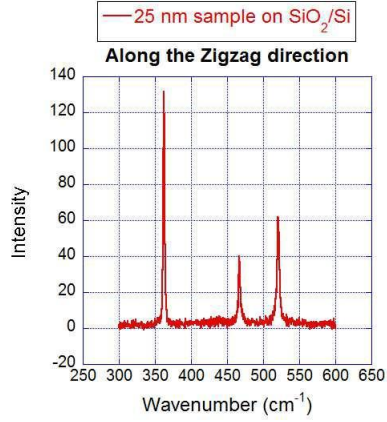
— 25 nm sample on SiO$_2$/Si

**Along the Zigzag direction**

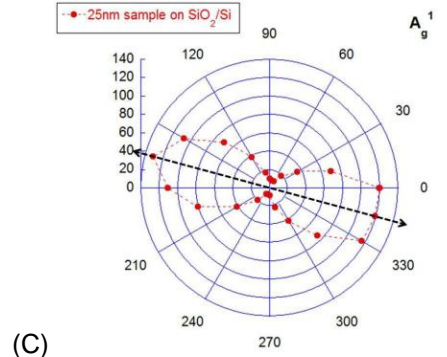
25nm sample on SiO$_2$/Si $\quad A_g^1$

(A)      (B)      (C)

| Optical image | Raman spectrum | Polar diagram of $A_g^2$ mode |
|---|---|---|

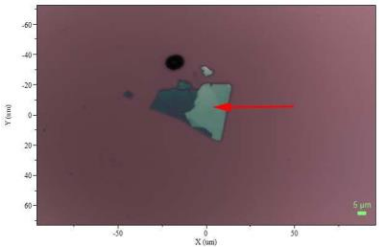

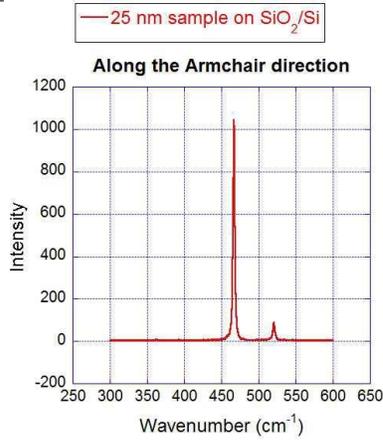
— 25 nm sample on SiO$_2$/Si

**Along the Armchair direction**

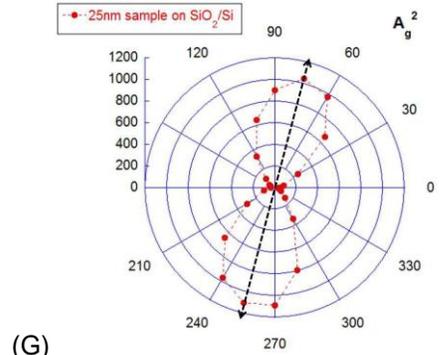
25nm sample on SiO$_2$/Si $\quad A_g^2$

(E)      (F)      (G)

| Optical image | Raman spectrum | Polar diagram of $A_g^1$ mode |
|---|---|---|

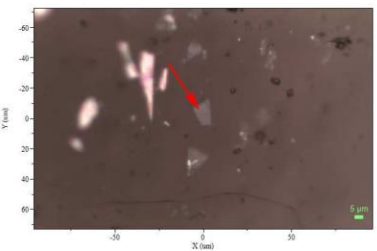

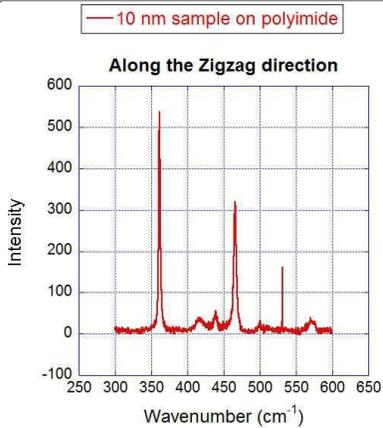
— 10 nm sample on polyimide

**Along the Zigzag direction**

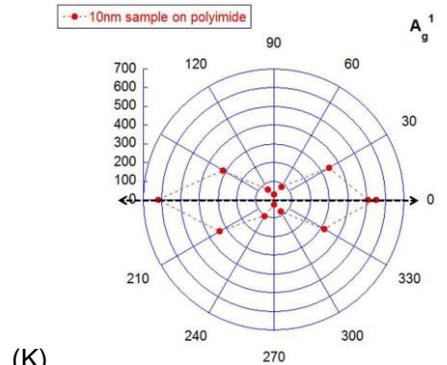
10nm sample on polyimide $\quad A_g^1$

(I)      (J)      (K)

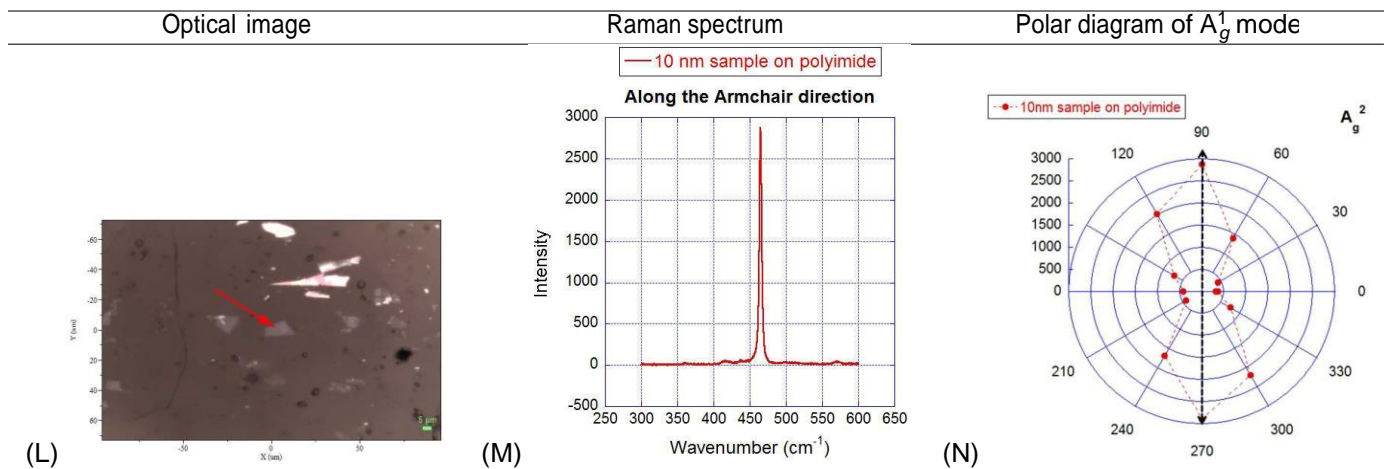

| Optical image | Raman spectrum | Polar diagram of $A_g^1$ mode |
| --- | --- | --- |

(L)      (M)      (N)

**Figure 1.** (A)-(G) are the results of 25-nm-thick BP on a SiO$_2$/Si substrate; (A) and (E) are the optical images of the BP sample, where the arrows are the zigzag directions of the sample; (B) and (F) are the corresponding Raman spectra; (C) and (G) are the polar diagrams of $A_g^1$ and $A_g^2$ modes; Fig.(I)-(N) are the results of 10-nm-thick BP on a polyimide substrate, with the optical images, Raman spectra and polar diagrams.

| No. | SG | PG | $\{E\|0\}$ | $\{C_{2x(z=\frac{1}{4})}\|\gamma_x^{(1)}\}$ | $\{C_{2y(x=z=\frac{1}{4})}\|0\}$ | $\{C_{2x}\|0\}$ | $\{i\|0\}$ | $\{\sigma_{xy}\|0\}$ | $\{\sigma_{xz}\|\gamma_{xz}^{(2)}\}$ | $\{\sigma_{yz(x=\frac{1}{4})}\|\gamma_z^{(3)}\}$ | Basis |
|---|---|---|---|---|---|---|---|---|---|---|---|
| 1. | $\Gamma_2^-$ | $B_{1u}$ | 1 | 1 | -1 | -1 | -1 | 1 | 1 | -1 | x |
| 2. | $\Gamma_4^-$ | $B_{3u}$ | 1 | -1 | -1 | 1 | -1 | -1 | 1 | 1 | z |
| 3. | $\Gamma_4^+$ | $B_{3g}$ | 1 | -1 | -1 | 1 | 1 | 1 | -1 | -1 | xy |
| 4. | $\Gamma_3^-$ | $B_{2u}$ | 1 | -1 | 1 | -1 | -1 | 1 | -1 | 1 | y |

**Table 1.** Character table with space group (SG) tables to point group (PG) for the $\Gamma$ point of black phosphorus with N odd number of layers with AB stacking[17,25]; according to Ref.[17], $B_{1u}$ is the representation for the armchair-polarized light and $B_{3u}$ for the zigzag-polarized light, and the satisfaction of transition from $B_{3g}$ to $B_{2u}$ (Excitation energy $\approx$ 2.8 eV) indicates that the intensity can reach its maximum when excitation polarization is parallel to armchair direction of BP.

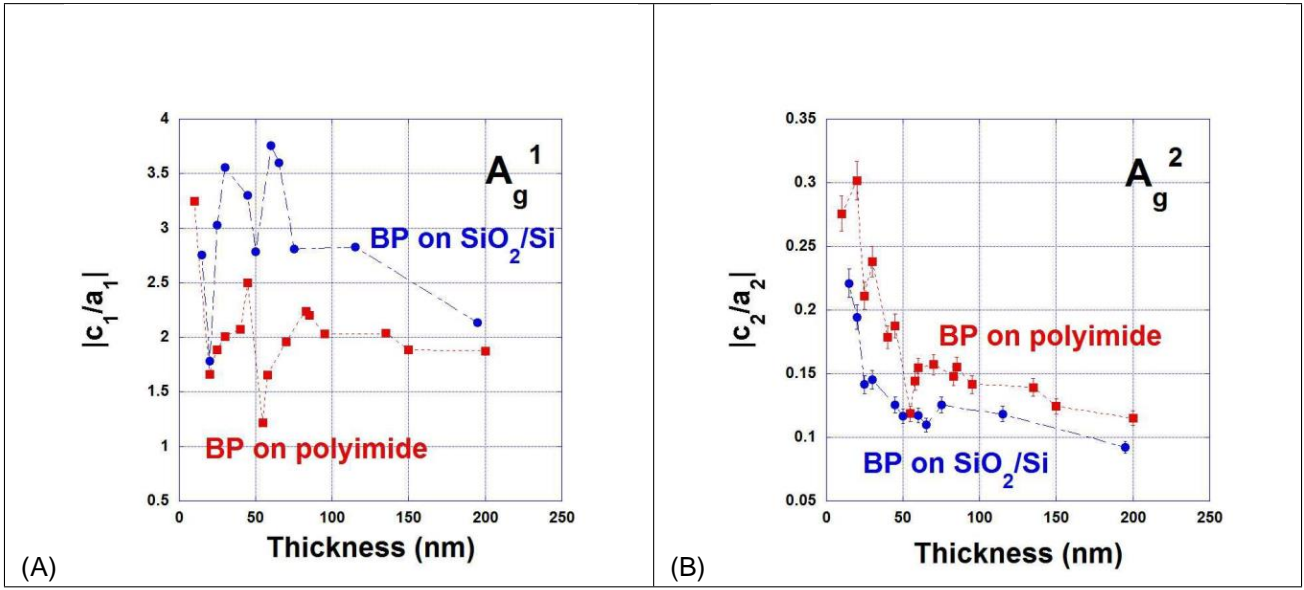

**Figure 2.** (A) and (B): Raman tensor ratio $|\frac{c}{a}|$ of $A_g^1$ and $A_g^2$ modes as functions of BP's thickness after removing the interference effects.

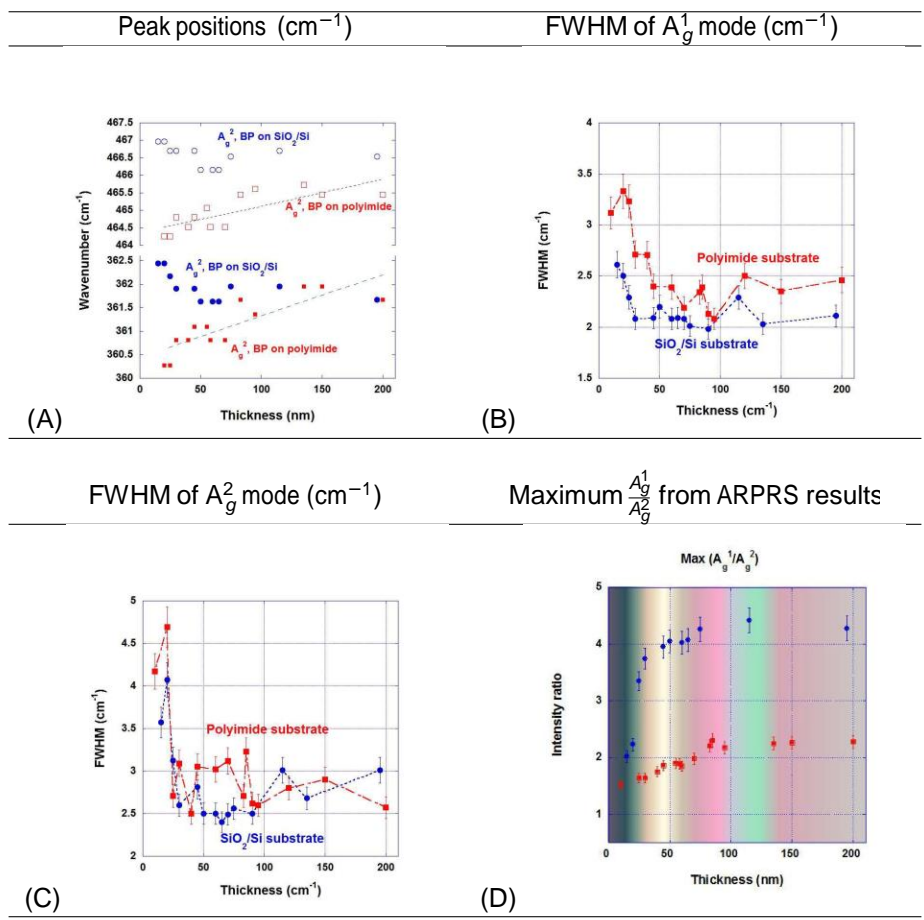

**Figure 3.** (A): peak positions of $A_g^1$ and $A_g^2$ modes as functions of thickness; (B)-(C): FWHM of $A_g^1$ and $A_g^2$ modes as function of thickness; (D): Maximum $\frac{A_g^1}{A_g^2}$ from ARPRS results is plotted as a function of BP's thickness. All data were collected with excitation polarization being parallel to zigzag direction of BP.

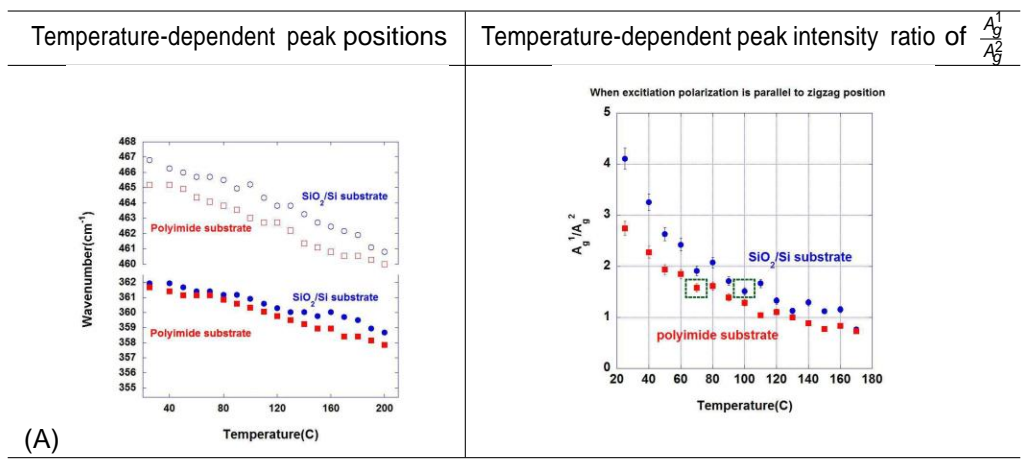

**Figure 4.** (A) Temperature-dependent peak positions of $A_g^1$ and $A_g^2$; (B) Temperature-dependent peak intensity ratio of $\frac{A_g^1}{A_g^2}$ All data were collected from in-situ Raman measurement during the thermal heating experiment. The excitation polarization was parallel to the zigzag direction of BP.

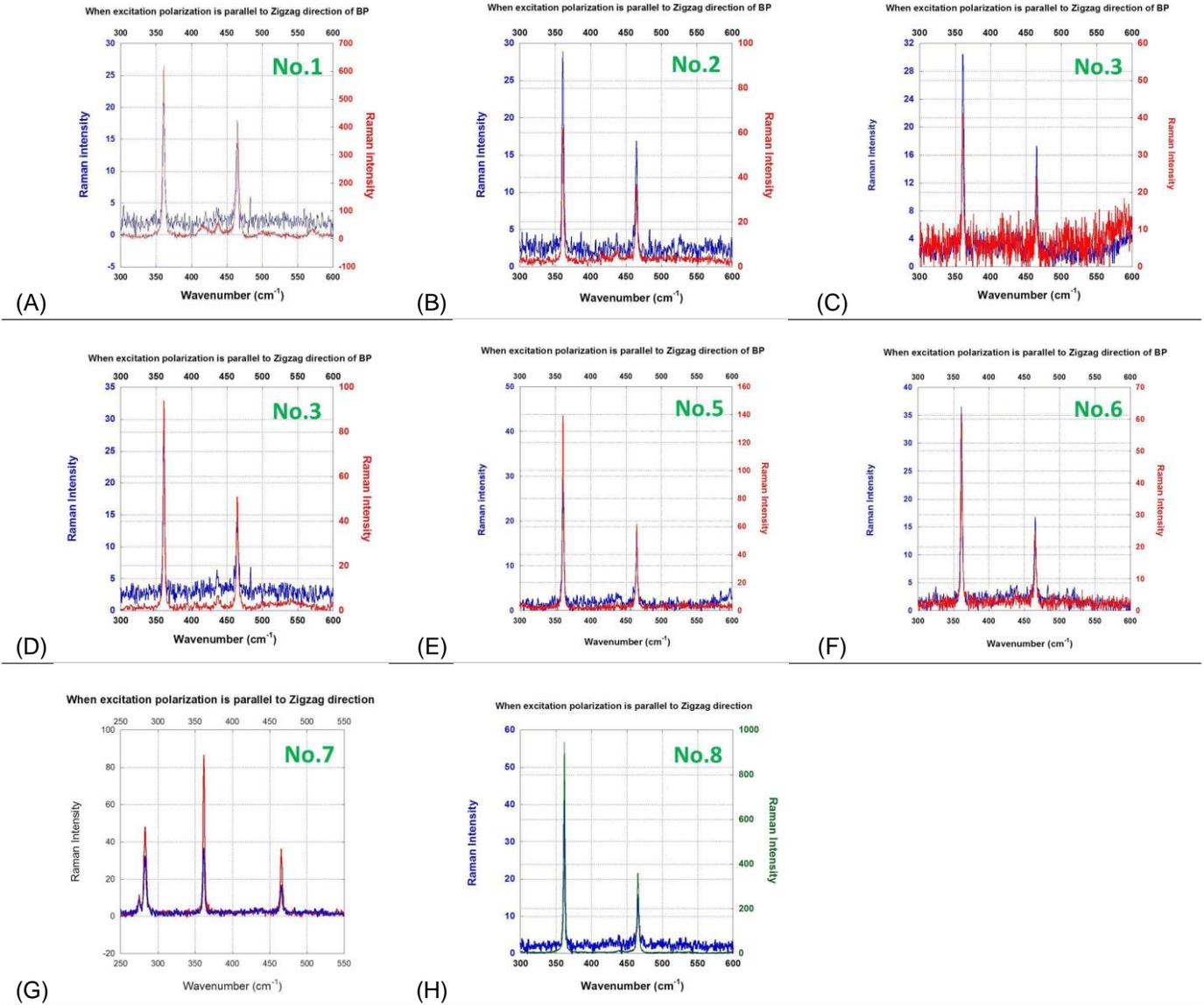

**Figure 5.** (A) - (G): according to intensity ratio of $\frac{A_g^1}{A_g^2}$ (from Figure 3 and Fig.(A)-(B) of Figure 4) and peak positions, we plotted the matched Raman spectra of BP sample on polyimide (red line) from ARPRS results and bulk BP on $SiO_2$/Si substrate (blue line) from in-situ thermal heating experiment;(H): by increasing the laser power by 10 times, the Raman spectrum (green) was plotted with the matched one (blue line) from in-situ thermal heating experiment;

**PART I. Laser heating effect in BP on polyimide substrate.**

| | | | | | | | |
|---|---|---|---|---|---|---|---|
| **BP on polyimide substrate** | | | | | | | |
| No. | Thickness (nm) | | Max($\frac{A_g^1}{A_g^2}$) | Peak position (cm$^{-1}$) | FWHM (cm$^{-1}$) | Peak position (cm$^{-1}$) | FWHM (cm$^{-1}$) |
| | | Prepared by: | | $A_g^1$ | $A_g^1$ | $A_g^2$ | $A_g^2$ |
| 1. | 10 | Sublimation | 1.46 | 360.81 | 3.12 | 465.07 | 4.17 |
| 2. | 30 | Sublimation | 1.65 | 360.81 | 2.69 | 464.8 | 3.05 |
| 3. | 40 | Sublimation | 1.75 | 360.54 | 2.46 | 464.8 | 2.42 |
| 4. | 45 | Exfoliation | 1.86 | 360.81 | 2.43 | 465.07 | 2.83 |
| 5. | 55 | Exfoliation | 1.9 | 361.09 | 2.11 | 465.07 | 2.5 |
| 6. | 83 | Exfoliation | 2.21 | 361.67 | 2.29 | 465.45 | 2.52 |
| 7. | 150 | Exfoliation | 2.26 | 361.67 | 2.35 | 465.72 | 2.9 |

| | | | | | | |
|---|---|---|---|---|---|---|
| **Thermal ramp Raman measurement: bulk BP on SiO$_2$/Si substrate: thickness >200nm, prepared by exfoliation** | | | | | | |
| | | Max($\frac{A_g^1}{A_g^2}$) | Peak position (cm$^{-1}$) | FWHM (cm$^{-1}$) | Peak position (cm$^{-1}$) | FWHM (cm$^{-1}$) |
| No. | Temperature (°C) | | $A_g^1$ | $A_g^1$ | $A_g^2$ | $A_g^2$ |
| 1. | 100 | 1.44 | 360.58 | 3.2 | 464.63 | 4.01 |
| 2. | 90 | 1.72 | 360.86 | 2.63 | 464.63 | 3.12 |
| 3. | 80 | 1.83 | 360.86 | 2.53 | 465.18 | 2.6 |
| 4. | 90 | 1.89 | 360.86 | 2.65 | 464.9 | 3.21 |
| 5. | 80 | 1.83 | 360.86 | 2.53 | 465.18 | 2.6 |
| 6. | 60 | 2.18 | 361.67 | 2.21 | 465.72 | 2.3 |
| 7. | 60 | 2.18 | 361.67 | 2.21 | 465.72 | 2.3 |

**Table 2.** Summary of intensity ratio of $\frac{A_g^1}{A_g^2}$, peak positions and full width at half maximum (FWMH) of $A_g^1$ and $A_g^2$ modes from Raman spectra listed in Fig.(A)-(I) of Figure 5.

**PART II. Laser heating effect in BP on SiO$_2$/Si substrate.**

| | | | | | | |
|---|---|---|---|---|---|---|
| **Bulk BP prepared by exfoliation on SiO$_2$/Si substrate:** | | | | | | |
| | | Max($\frac{A_g^1}{A_g^2}$) | Peak position (cm$^{-1}$) | FWHM (cm$^{-1}$) | Peak position (cm$^{-1}$) | (cm$^{-1}$) |
| No. | Laser power (mW) | | $A_g^1$ | $A_g^1$ | $A_g^2$ | $A_g^2$ |
| 8. | 2 | 2.63 | 361.67 | 2.24 | 465.72 | 2.8 |
| **Thermal ramp Raman measurement: bulk BP on SiO$_2$/Si substrate: thickness >200nm, prepared by exfoliation** | | | | | | |
| | | Max($\frac{A_g^1}{A_g^2}$) | Peak position (cm$^{-1}$) | FWHM (cm$^{-1}$) | Peak position (cm$^{-1}$) | FWHM (cm$^{-1}$) |
| No. | Temperature (°C) | | $A_g^1$ | $A_g^1$ | $A_g^2$ | $A_g^2$ |
| 8. | 50 | 2.64 | 361.67 | 2.31 | 465.99 | 2.94 |

**Table 3.** Summary of intensity ratio of $\frac{A_g^1}{A_g^2}$, peak positions and full width at half maximum (FWMH) of $A_g^1$ and $A_g^2$ modes from Raman spectra listed in Fig.(H) of Figure 5.

**All Raman spectra were taken at room temperature**

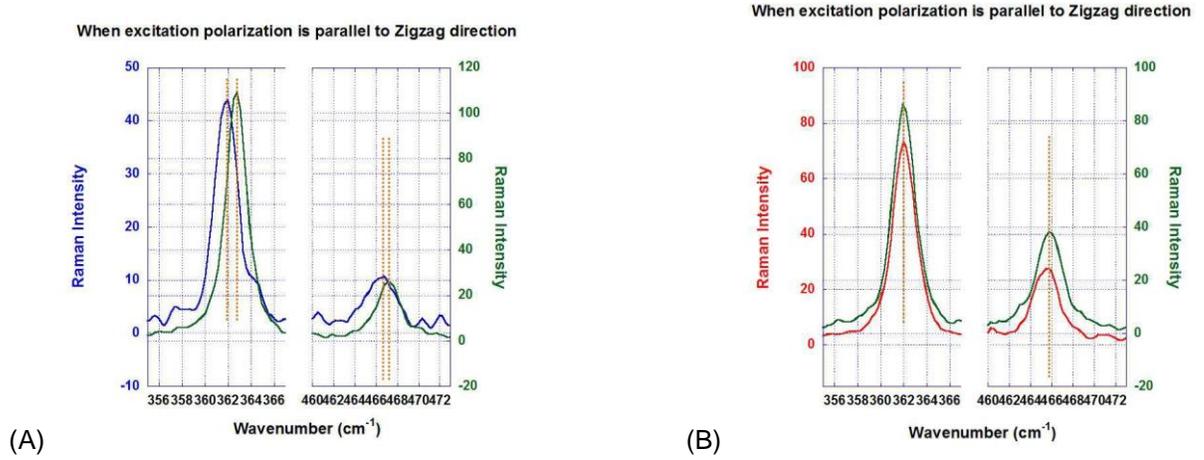

(A)                                          (B)

Figure 6. (A): Typical Raman spectra of the thick BP sample on SiO₂/Si substrate before (blue line) and after the thermal heating experiment (green line); (B): Typical Raman spectra of the thick BP sample on polyimide substrate before (red line) and after the thermal heating experiment (green line); the peak positions, FWHM and $\frac{A^1_g}{A^2_g}$ were listed.

| | SiO₂/Si substrate | | | | | polyimide substrate | | | | |
| | Peak position(cm⁻¹) | | FWMH(cm⁻¹) | | $\frac{A^1_g}{A^2_g}$ | Peak position(cm⁻¹) | | FWMH(cm⁻¹) | | $\frac{A^1_g}{A^2_g}$ |
| | $A^1_g$ | $A^2_g$ | $A^1_g$ | $A^2_g$ | | $A^1_g$ | $A^2_g$ | $A^1_g$ | $A^2_g$ | |
|---|---|---|---|---|---|---|---|---|---|---|
| Before thermal heating | 361.95 | 466.81 | 2.37 | 3.33 | 4.11 | 361.95 | 465.72 | 2.43 | 2.81 | 2.58 |
| After thermal heating | 362.76 | 467.35 | 2.17 | 2.61 | 4.24 | 361.95 | 465.72 | 2.41 | 2.86 | 2.26 |

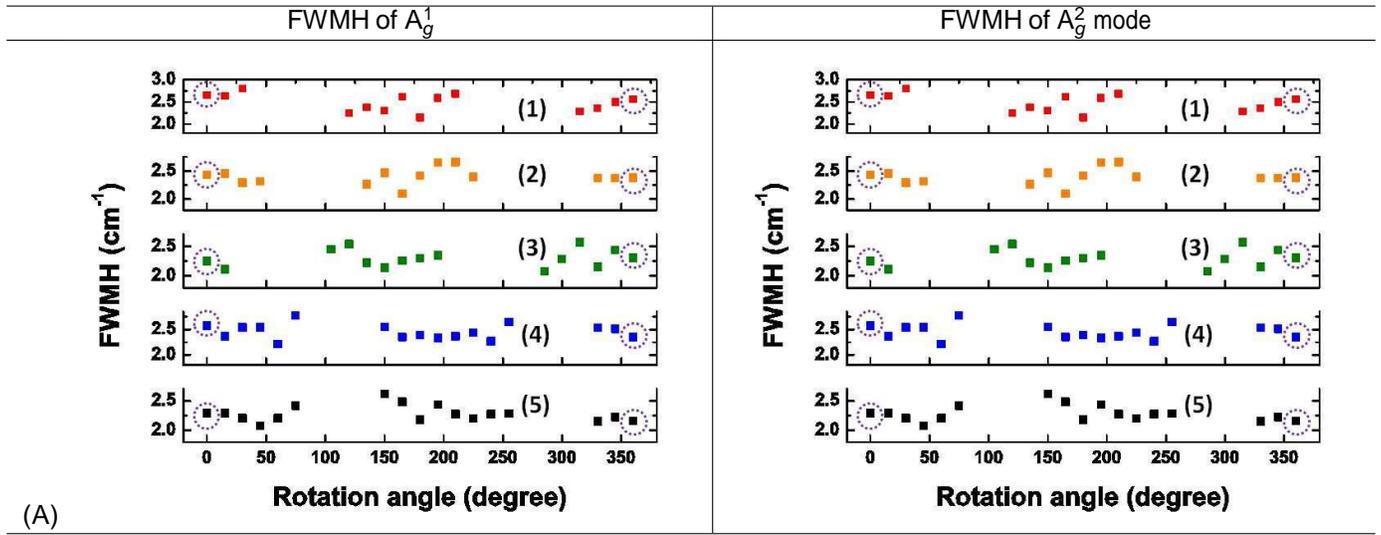

(A)

**Figure 7.** (1) - (5) are 15-nm, 20-nm, 25-nm, 45-nm and 50 nm-thick samples on $SiO_2$/Si substrates. (A) FWHM of $A^1_g$ mode as a function of rotation angle (time); (B) FWHM of $A^2_g$ mode as a function of rotation angle (time).

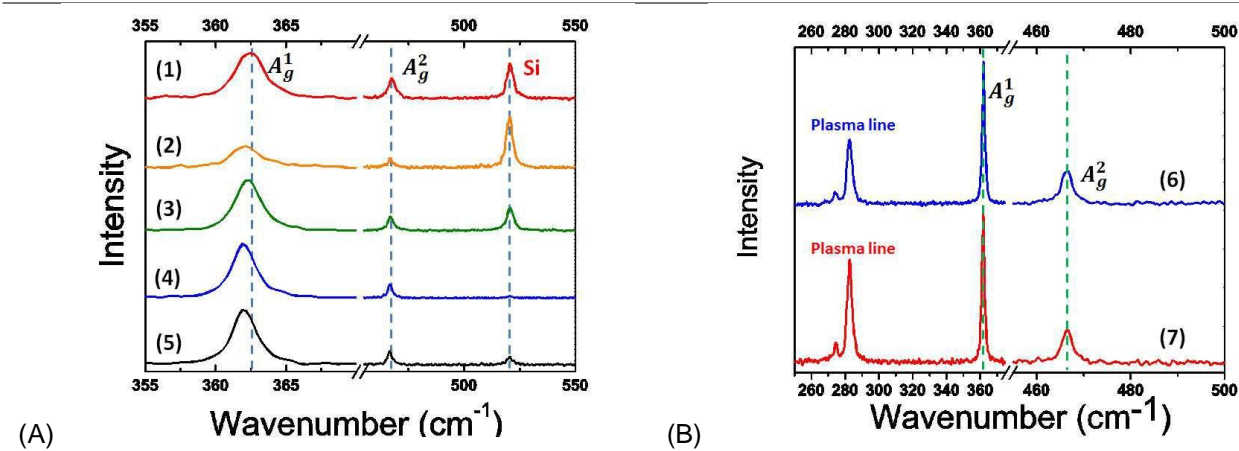

| Prepared by thermal sublimation | | | | Prepared by exfoliation | | | |
|---|---|---|---|---|---|---|---|
| No. | Thickness (nm) | Peak position(cm⁻¹) | | No. | Thickness (nm) | Peak position(cm⁻¹) | |
| | | $A_g^1$ | $A_g^2$ | | | $A_g^1$ | $A_g^2$ |
| 1. | 15 | 362.44 | 467.51 | 6. | 75 | 361.95 | 466.54 |
| 2. | 20 | 362.17 | 466.70 | 7. | 195 | 361.67 | 466.54 |
| 3. | 25 | 362.17 | 466.97 | | | | |
| 4. | 30 | 361.63 | 466.70 | | | | |
| 5. | 45 | 361.63 | 466.70 | | | | |

**Figure 8.** (A): Typical Raman spectra of 15 nm, 20 nm, 25 nm, 25 nm, 30 nm and 45 nm samples on SiO₂/Si substrates, all samples were prepared by thermal sublimation, here we used Si peak at 520.45 (cm⁻¹) for reference; Typical Raman spectra of 75 nm and 195 nm samples on SiO₂/Si substrates, both samples were prepared by mechanical exfoliation. Here we used the plasma line of 442 nm excitation at 282.68 (cm⁻¹) instead of Si peak for reference because 442 nm cannot penetrate the 195-nm-thick BP sample.

# Supporting information:

**Contents:**

1. Investigation in Laser heating effect on BP samples on $SiO_2$/polyimide substrates.

2. AFM data of BP samples on $SiO_2$/Si and polyimide substrates before and after heating process.

3. Polar diagrams of $A_g^1$ and $A_g^2$ of BP samples with various thickness, on $SiO_2$/Si and polyimide substrates, acquired by ARPRS with 442 nm excitation.

4. Polar diagrams of $\frac{A_g^1}{A_g^2}$ and $\frac{A_g^2}{A_g^1}$ of BP samples with various thickness, on $SiO_2$/Si and polyimide substrates, acquired by ARPRS with 442 nm excitation.

5. Calculations of interference enhancement factor of as function of thickness.

6. Calculation of amplitudes of Raman tensors $|a|$ and $|c|$.

7. $A_g^1$ mode was too weak to be observable when excitation polarization was parallel to zigzag direction of BP.

## 1. Laser heating effect on SiO$_2$/Si and polyimide substrate

| Material | Thermal conductivity (W·m$^{-1}$K$^{-1}$) | Thermal expansion coefficient (ppm/K) |
|---|---|---|
| Black phosphorus (> 15 nm) | 40 (Zigzag) / 20 (Armchair)[23] | 90.3 (Zigzag) / 93.2 (Armchair)[38] |
| Polyimide[39] | 0.12[39] | 20 (30 − 100°C) / 32 (100 − 200°C) / 40 (200 °C-300 °C) |
| Si (100) | 130-150[40,41,42] | 2.6[43,44] |

**Table S1** Parameters of materials used in this study.

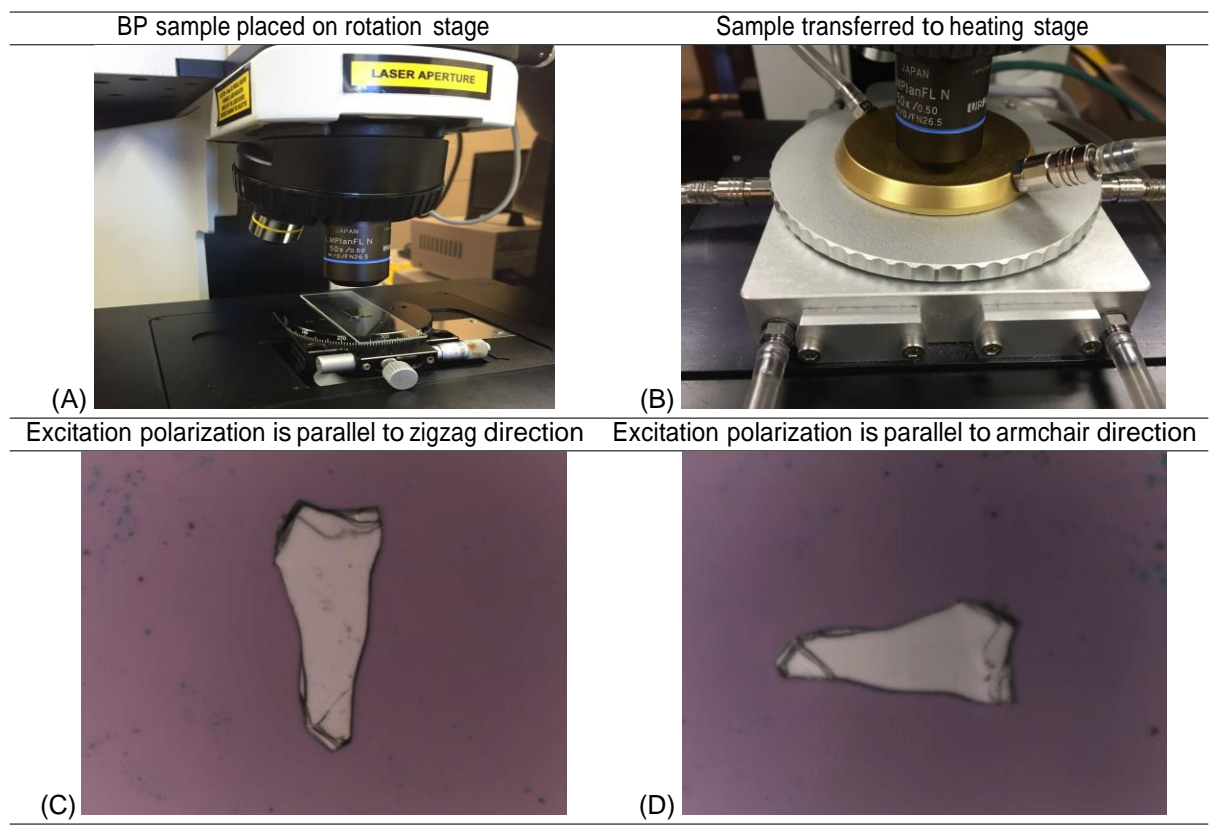

**Table S2** Optical images of experimental setup: (A) A typical BP sample on SiO$_2$/Si subtrate placed on rotation stage for ARPRS study; (B) A BP sample placed in heating stage for high-temperature Raman spectroscopy; (C) and (D), illustrations of the BP in heating stage: when the excitation polarization is parallel to the Zigzag/armchair direction of BP.

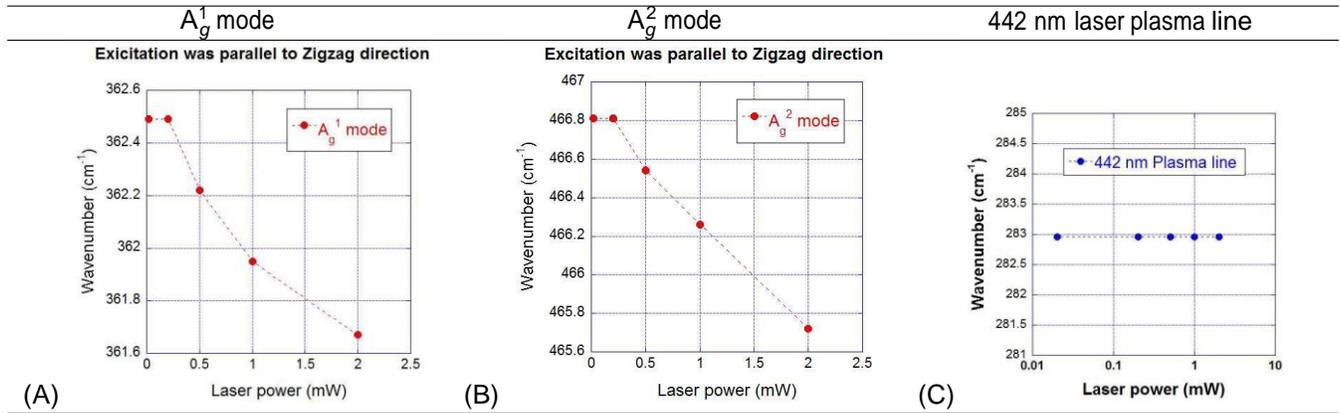

| $A_g^1$ mode | $A_g^2$ mode | 442 nm laser plasma line |
|---|---|---|

**Table S3** Fig.(A)-(B): Measured peak positions of BP's $A_g^1$ and $A_g^2$ modes as functions of filter used to reduce the laser power; (C): the peak position of 442 nm plasma line is plotted as a function of added filter, which is used for referencing the shifts of peak position .

In Table S2 , optical images of the experimental setup are listed: Fig.(A) denotes a typical BP sample placed on rotation angle while Fig.(B) shows a BP sample placed in heating stage.

In Table S3, we investigated laser-power dependent peak positions of BP samples on SiO$_2$/Si substrate with excitation polarization being parallel to zigzag direction of BP. By using peak position of 442 nm plasma line at 283 cm$^{-1}$[45] as a reference, we demonstrated that, the peak positions started blue-shifting with laser power greater than 0.2 mW, relating to filter D=1. Thus to avoid the laser heating effect, we chose the laser power of 0.2 mW in whole study.

After that, in Fig.(A) of Table S4, investigation on laser-power dependent peak positions of BP samples on polyimide substrate, with excitation polarization being parallel to zigzag direction of BP, exhibited even using D=1 filter (laser power being reduced to 0.2 mW), the laser heating effect still caused blue shifts of BP's peak positions, which was also supported by the decreasing peak intensity ratio of $\frac{A_g^1}{A_g^2}$ showed in Fig.(B) of Table S2. In addition, thermal conductivity of Si was reported to be 130 - 150 W/m·K[40,41,42] , while that of polyimide was 0.12 W/m·K.[39] This also could help to explain that the heat introduced by laser dissipates was much harder to dissipate on polyimide than on SiO$_2$/Si, which resulted in the promotion of temperature.

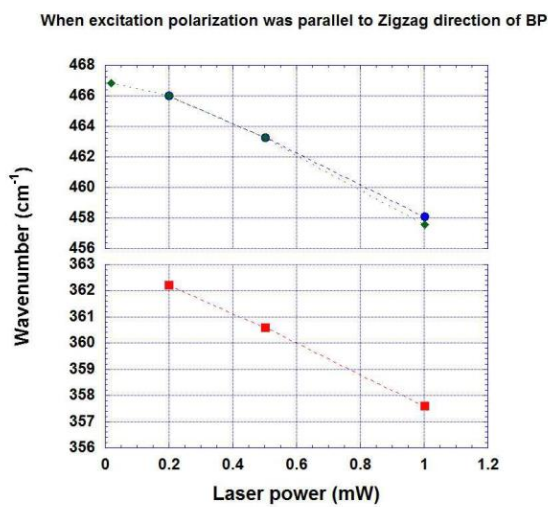
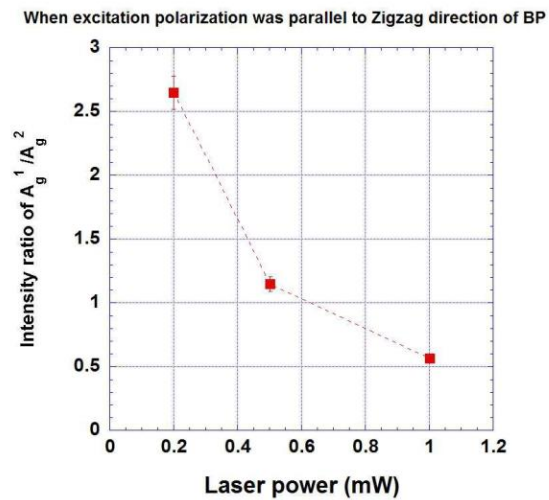

**Table S4** (A) Measured peak positions of BP's $A_g^1$ and $A_g^2$ modes as functions of laser power (a typical thick ($>200nm$) BP flake on polyimide substrate): the red cubes and blue dots represent peak positions of $A_g^1$ and $A_g^2$ modes when the excitation polarization was parallel to zigzag direction, and the green diamonds denote the peak positions of $A_g^2$ when the excitation polarization was parallel to armchair direction ; (B) The peak intensity ratio of $\frac{A_g^1}{A_g^2}$ of a typical thick ($>200nm$) BP flake on polyimide substrate, acquired from measured data when the excitation polarization was parallel to BP's zigzag direction, is plotted as a function of laser power.

**2. AFM data of BP samples on SiO$_2$/Si and polyimide substrates before and after heating process**

| No. | Thickness | Optical image | AFM morphology | AFM profile |
|-----|-----------|---------------|----------------|-------------|
| 1. | 15 nm |  |  |  |
| 2. | 20 nm |  |  |  |
| 3. | 25 nm |  |  |  |
| 4. | 30 nm |  |  |  |
| 5. | 45 nm |  |  |  |
| 6. | 50 nm |  |  |  |
| 7. | 60 nm |  |  |  |
| 8. | 65 nm |  |  |  |
| 9. | 75 nm | 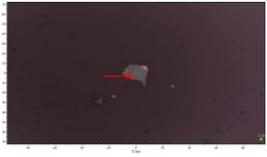 | 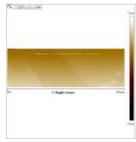 | 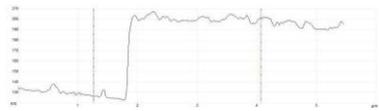 |
| 10. | 115 nm | 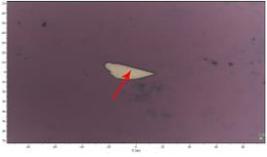 | 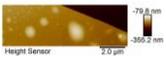 | 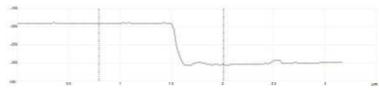 |
| 11. | 195 nm | 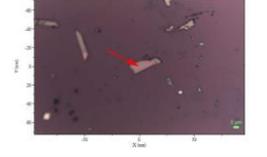 | 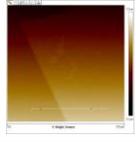 | 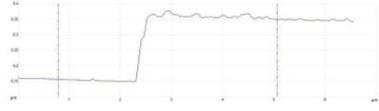 |

**Table S5** Optical images, AFM images and thickness profiles of 11 BP Samples on SiO$_2$/Si substrate. No.1-5 were samples prepared by sublimation thinning; No.6-12 were samples prepared by exfoliation.

| No. | Thickness | Optical image | AFM morphology | AFM profile |
|-----|-----------|---------------|----------------|-------------|
| 1. | 10 nm | | | |
| 2. | 20 nm | | | |
| 3. | 25 nm | | | |
| 4. | 30 nm | | | |
| 5. | 40 nm | | | |
| 6. | 45 nm | | | |
| 7. | 55 nm | | | |
| 8. | 58 nm | | | |
| 9. | 60 nm | | | |
| 10. | 70 nm | | | |
| 11. | 83 nm | | | |
| 12. | 85 nm | | | |
| 13. | 95 nm | | | |
| 14. | 135 nm | | | |
| 15. | 150 nm | | | |
| 16. | 200 nm | | | |

**Table S6** Optical images, AFM images and thickness profiles of 16 BP Samples on polyimide substrate. No.1-5, 8, 10 and 12 were samples prepared by sublimation thinning; No.6-7, 9, 11 and 13-16 were samples prepared by exfoliation.

**3. Polar diagrams of $A_g^1$ and $A_g^2$ of BP samples with various thickness, on SiO$_2$/Si and polyimide substrates, acquired by ARPRS with 442 nm excitation.**

| No. | Thickness (nm) | Raman profile: $A_g^1$ mode appears to be maximum | Polar diagram of $A_g^1$ mode | Raman profile: $A_g^2$ mode appears to be maximum | Polar diagram of $A_g^2$ mode |
|---|---|---|---|---|---|
| 1. | 15 |  |  |  |  |
| 2. | 20 |  |  |  |  |
| 3. | 25 |  |  |  |  |
| 4. | 30 |  |  |  |  |
| 5. | 45 |  |  |  |  |

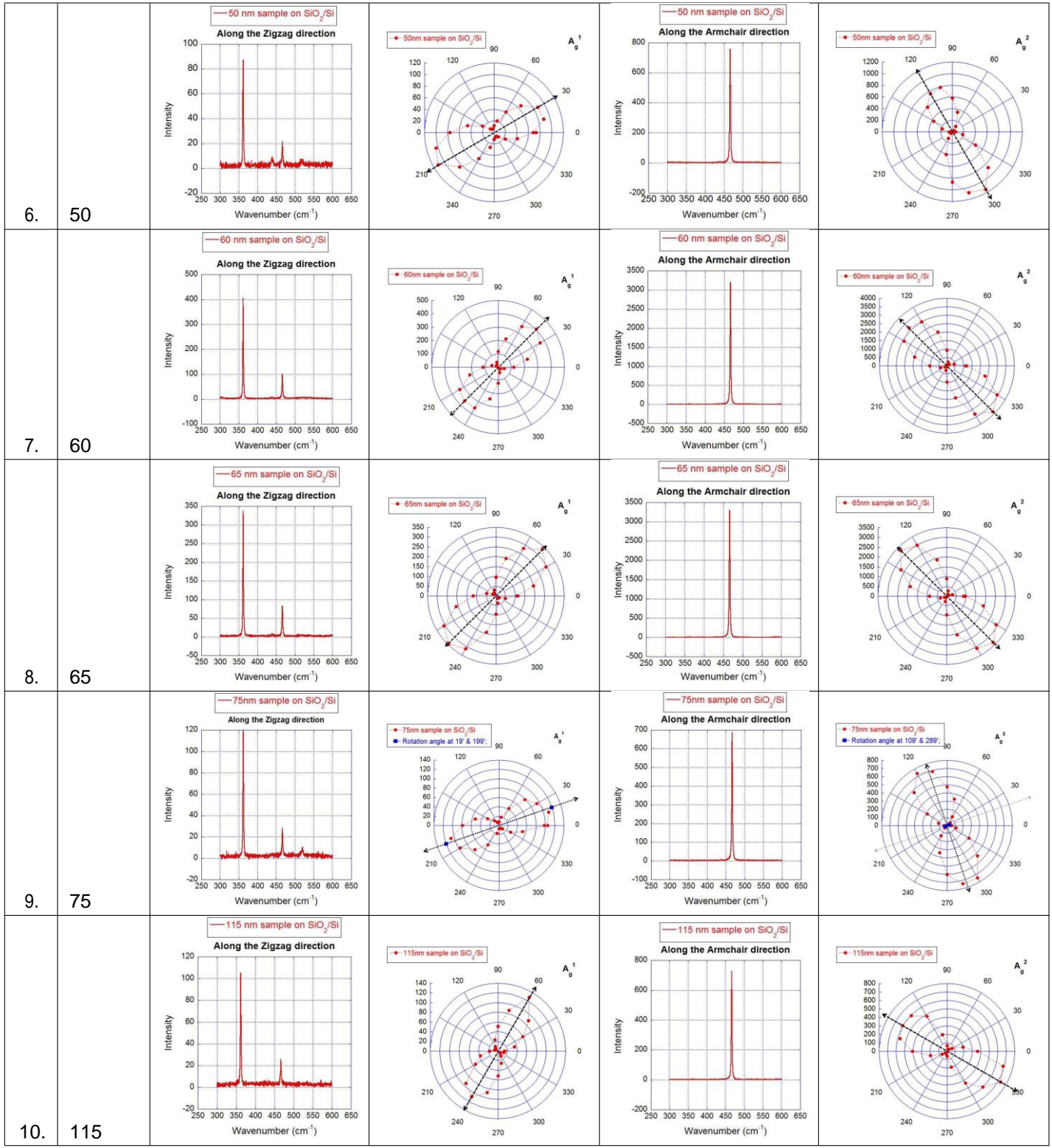

| 6. | 50 | | | | |
|----|----|----|----|----|----|
| 7. | 60 | | | | |
| 8. | 65 | | | | |
| 9. | 75 | | | | |
| 10. | 115 | | | | |

| | | |
|---|---|---|
| 11. | 195 | 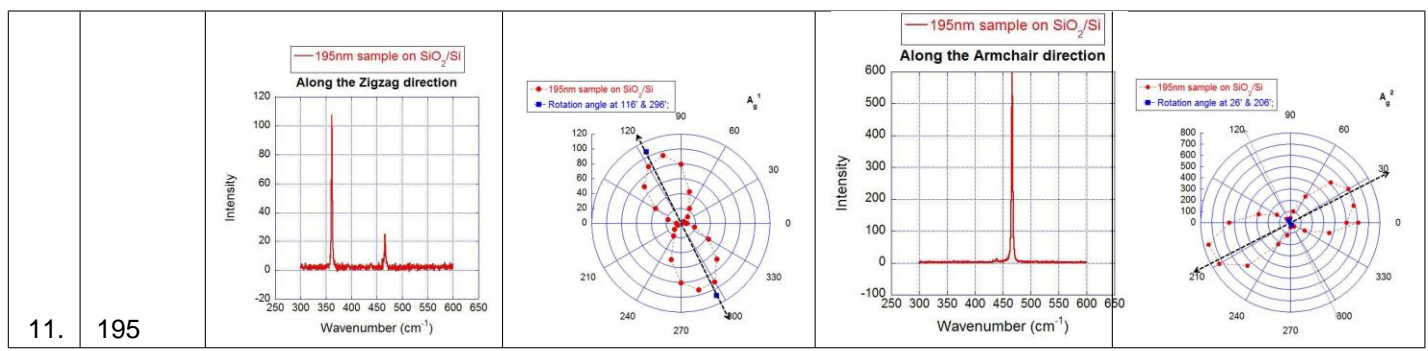 |

**Table S7** Polar diagrams of peak intensity ratios of $A^1_g$ over $A^2_g$ of 11 BP samples on $SiO_2$/Si substrates acquired by Angle-resolved Polarized Raman Spectroscopy (ARPRS): peak intensities of $A^1_g$ and $A^2_g$ modes as functions of rotation angles.

| No. | Thickness (nm) | Raman profile: $A_g^1$ mode appears to be maximum | Polar diagram of $A_g^1$ mode | Raman profile: $A_g^2$ mode appears to be maximum | Polar diagram of $A_g^2$ mode |
|---|---|---|---|---|---|
| 1. | 10 |  |  |  |  |
| 2. | 20 |  |  |  |  |
| 3. | 25 |  |  |  |  |
| 4. | 30 |  |  |  |  |
| 5. | 40 |  |  |  |  |

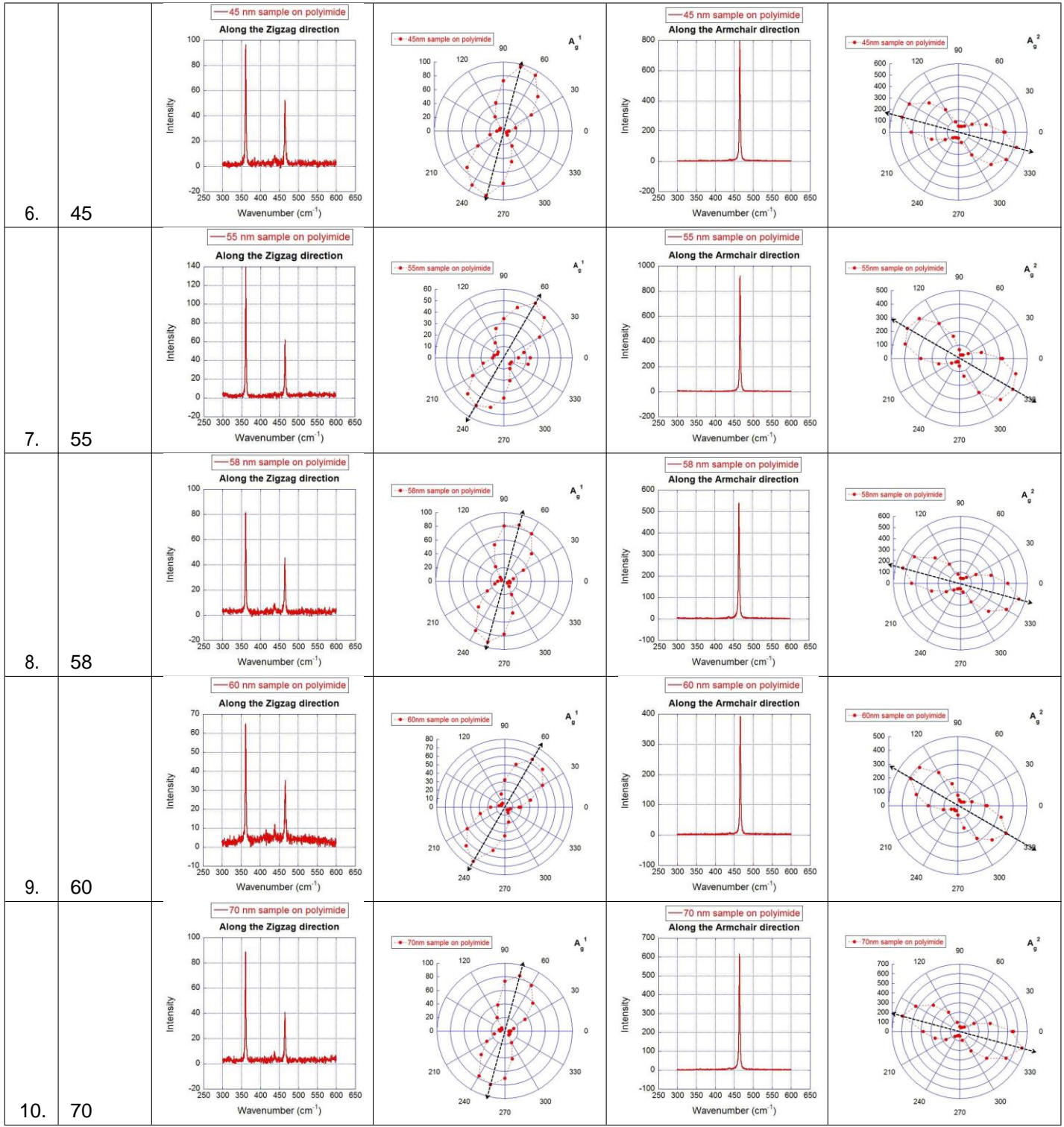

| | | | | | | |
|---|---|---|---|---|---|---|
| 6. | 45 | | | | | |
| 7. | 55 | | | | | |
| 8. | 58 | | | | | |
| 9. | 60 | | | | | |
| 10. | 70 | | | | | |

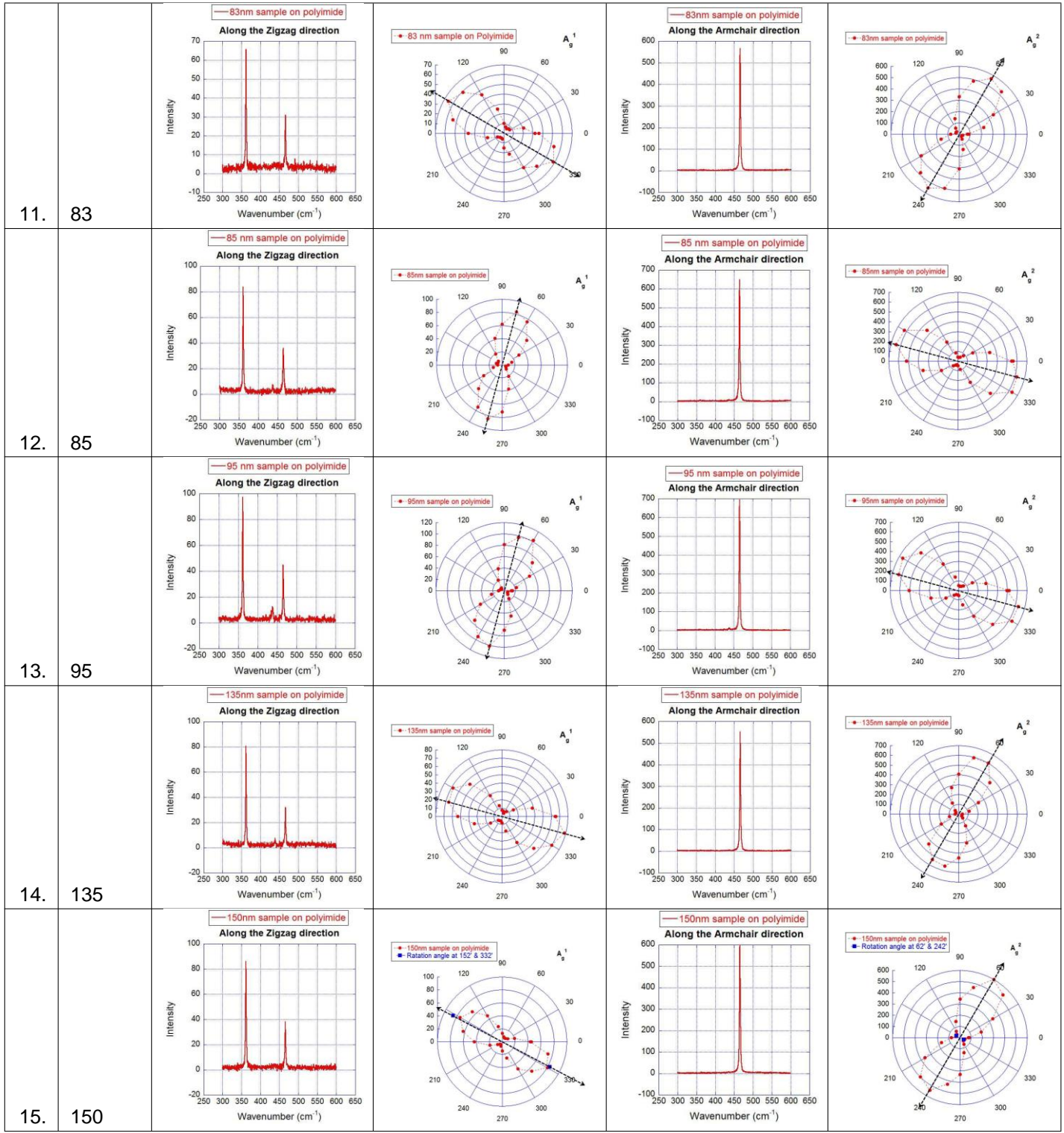

| 11. | 83 | | | | |
| 12. | 85 | | | | |
| 13. | 95 | | | | |
| 14. | 135 | | | | |
| 15. | 150 | | | | |

| 16. | 200 | 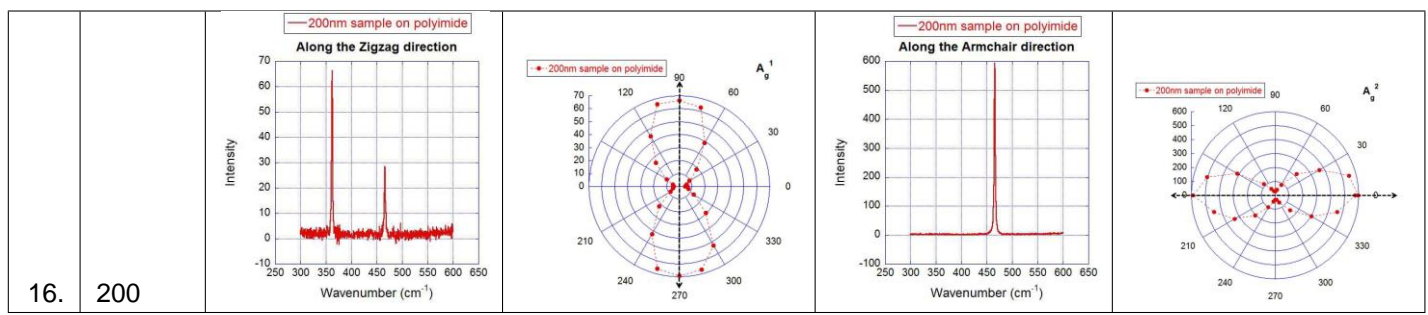 |

Table S8 Polar diagrams of 16 BP samples on polyimide substrates acquired by Angle-resolved Polarized Raman Spectroscopy (ARPRS): peak intensities of $A_g^1$ and $A_g^2$ modes as functions of rotation angles.

**4.** Polar diagrams of $\frac{A_g^1}{A_g^3}$ and $\frac{A_g^2}{A_g^3}$ of BP samples with various thickness, on SiO$_2$/Si and polyimide substrates, acquired by ARPRS, using 442 nm laser under parallel configuration. To mention that, when incident laser was along armchair direction, $A_g^1$ mode was invisible in most samples. Here in convenience of calculating the ratios between $A_g^1$ and $A_g^2$, for BP samples on SiO$_2$/Si and polyimide substrates, we simply used the substrate intensity to present that of $A_g^1$

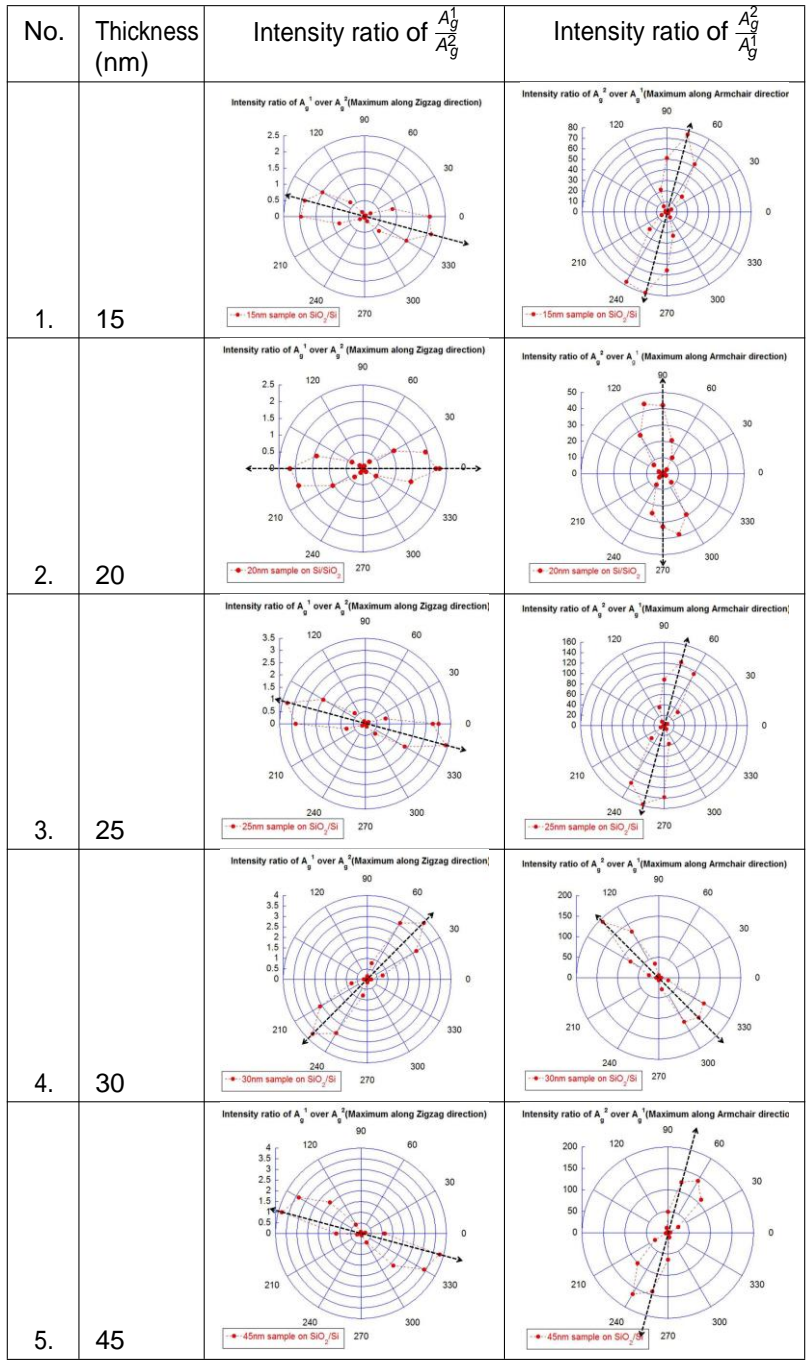

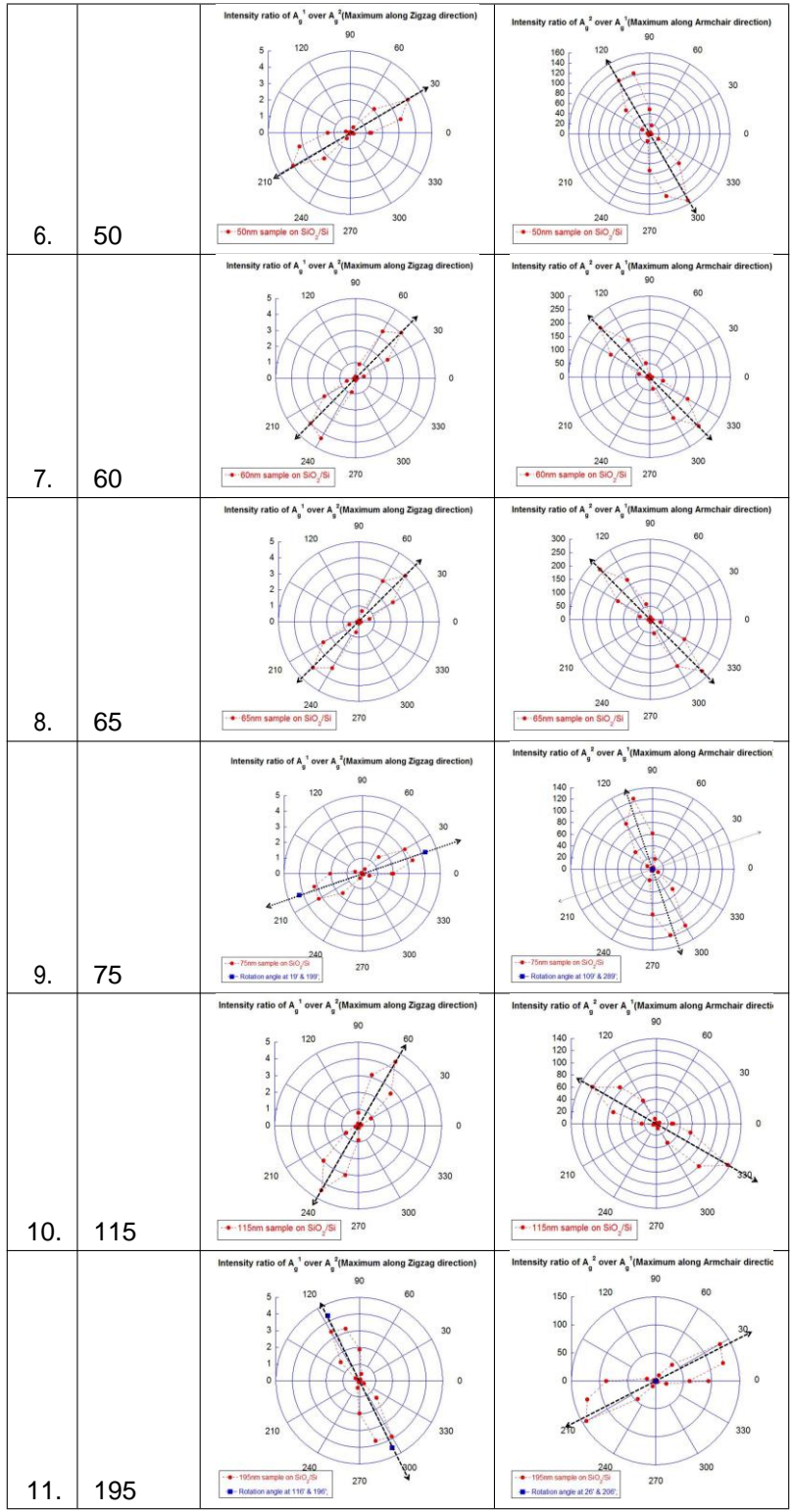

| 6. | 50 |
| 7. | 60 |
| 8. | 65 |
| 9. | 75 |
| 10. | 115 |
| 11. | 195 |

**Table S9** Polar diagrams of 11 BP samples on $SiO_2$/Si substrates acquired by Angle-resolved Polarized Raman Spectroscopy (ARPRS): peak intensity of $\frac{A_g^1}{A_g^2}$ and $\frac{A_g^2}{A_g^1}$ as functions of rotation angles.

| No. | Thickness (nm) | Intensity ratio of $\frac{A_g^1}{A_g^2}$ | Intensity ratio of $\frac{A_g^2}{A_g^1}$ |
|---|---|---|---|
| 1. | 10 |  |  |
| 2. | 20 |  |  |
| 3. | 25 |  |  |
| 4. | 30 |  |  |
| 5. | 40 |  |  |
| 6. | 45 |  |  |
| 7. | 55 | 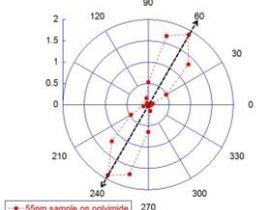 | 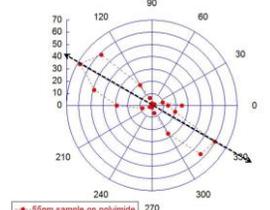 |
| 8. | 58 | 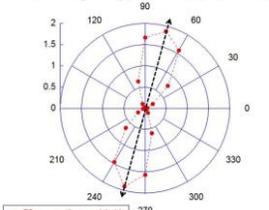 | 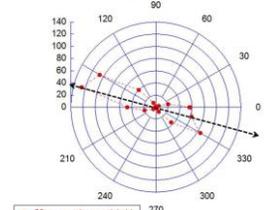 |
| 9. | 60 | 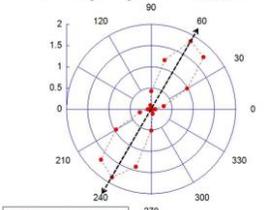 | 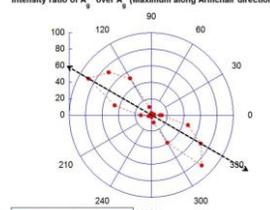 |
| 10. | 70 | 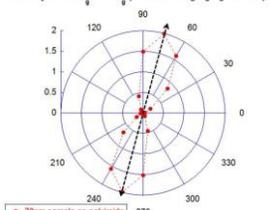 | 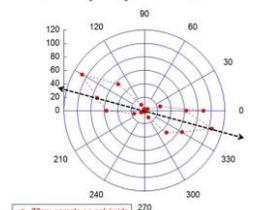 |
| 11. | 80 | 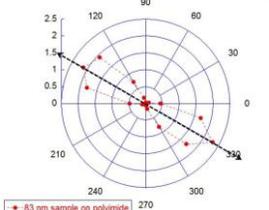 | 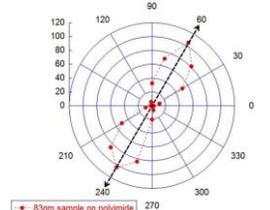 |
| 12. | 85 | 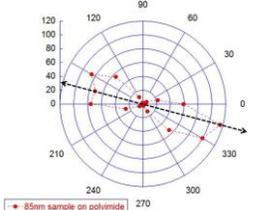 | 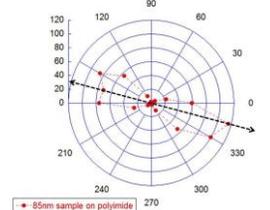 |

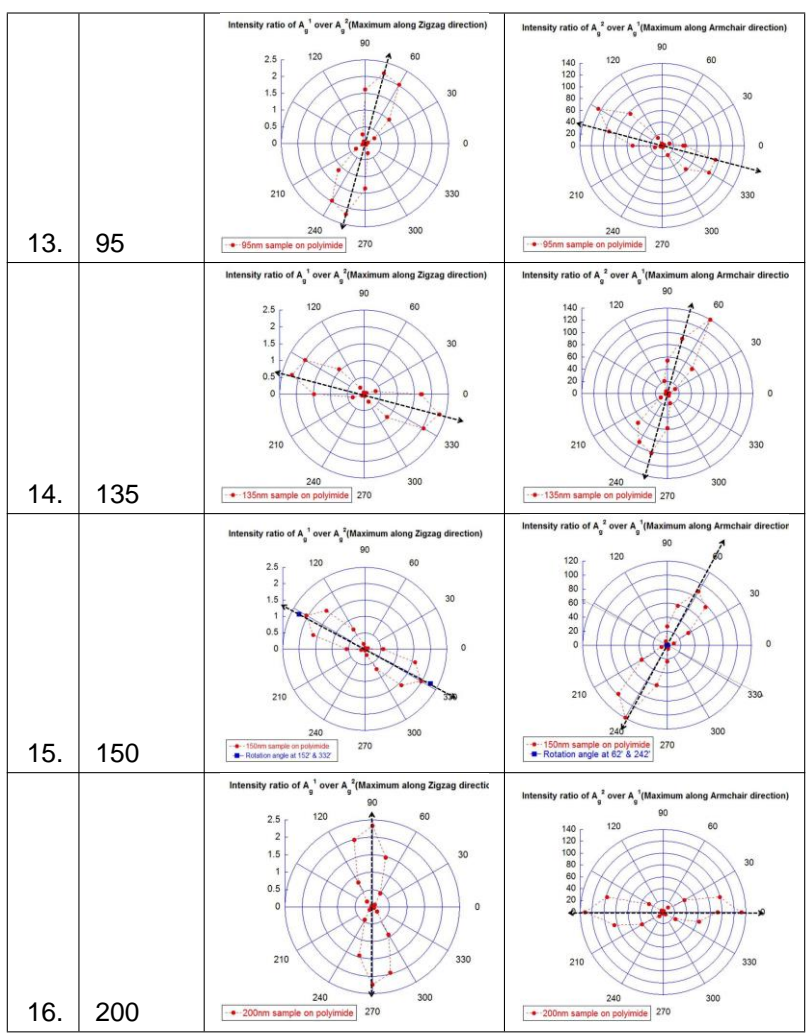

| | |
|---|---|
| 13. | 95 |
| 14. | 135 |
| 15. | 150 |
| 16. | 200 |

**Table S10** Polar diagrams of 16 BP samples on polyimide substrates acquired by Angle-resolved Polarized Raman Spectroscopy (ARPRS): peak intensity of $\frac{A_g^1}{A_g^2}$ and $\frac{A_g^2}{A_g^1}$ as functions of rotation angles.

**5. Calculations of interference enhancement factor of as function of thickness.**

Unlike the isotropic refractive indices of Graphene[46] and MoS$_2$,[47] BP shows different refractive indices along zigzag and armchair directions.[16,17] And here we followed their values[16] in our calculations: the total interference enhancement factor is then given by:

$$F(x) = \int_0^{d_1} |F_{in}(x) \times F_{out}(x)|^2 dx; \tag{1}$$

Here $F_{in}$ and $F_{out}$ are the net interference enhancement factors of the incident and scattered light enhanced by multi-reflections at a specific position x of BP surface. $F_{in}$ can be written as

$$F_{in}(x) = t_{01} \frac{(1 + r_{12} r_{23} e^{-2i\beta_2^{in}}) e^{-2i\beta_x^{in}} + (r_{12} + r_{23} e^{-2i\beta_2^{in}}) e^{-i(2\beta_1^{in} - \beta_x^{in})}}{1 + r_{12} r_{23} e^{-2i\beta_2^{in}} + (r_{12} + r_{23} e^{-2i\beta_2^{in}}) r_{01} e^{-2i\beta_1^{in}}} \tag{2}$$

where $t_{ij} = \frac{2n_i}{n_i + n_j}$, and $r_{ij} = \frac{n_i - n_j}{n_i + n_j}$ are the Fresnel transmittance and reflectance coefficients at the interfaces of the $i$th and $j$th layer; the subscripts i,j are assigned following the order of air (0), BP (1), SiO$_2$/polyimide (2), and Si / air (3); $n_i$ is the complex refractive index of the $i$th layer; $\beta_x^{in}=2\pi x n_1/\lambda_{in}$ and $\beta_i^{in}=2\pi d_i n_i/\lambda_{in}$ are the phase terms; $d_i$ is the thickness of $i$th layer; $\lambda_{in}$ = 442 nm (the excitation wavelength).

$F_{out}$ can be written as:

$$F_{in} = t_{10} \frac{(1 + r_{12} r_{23} e^{-2i\beta_2^{out}}) e^{-2i\beta_x^{out}} + (r_{12} + r_{23} e^{-2i\beta_2^{out}}) e^{-i(2\beta_1^{out} - \beta_x^{out})}}{1 + r_{12} r_{23} e^{-2i\beta_2^{out}} + (r_{12} + r_{23} e^{-2i\beta_2^{out}}) r_{01} e^{-2i\beta_1^{out}}} \tag{3}$$

where $\beta_x^{out}=2\pi x n_1/\lambda_{out}$ and $\beta_i^{out}=2\pi d_i n_i/\lambda_{out}$ are the phase terms; $d_i$ is the thickness of $i$th layer; in our study, $\lambda_{out}$ denotes for the wavelength of BP's A$_g^1$ (or A$_g^2$) mode. Here to note that, following Ref.[16] which added a constant normalization factor $\boldsymbol{N}$ on the total $\boldsymbol{F}$ factor, no normalization factor was applied in our calculation. Both were correct in that only the thickness of BP sample varied in our study with fixed excitation wavelength of 442 nm. The variation of interference effect was considered as a function of the sample thickness only. For the refractive indexes of BP, SiO$_2$ and Si, we followed previous work[16]; for that of 100 $\mu$m polyimide , we simply use the value of 1.79 (measured using Sodium D line: 590 nm).[39] The calculation results are lsited in Table S11.

| No | substrate | Enhancement factor | Enhancement ratio of $\frac{F_{Zigzag}}{F_{Armchair}}$ |
|---|---|---|---|
| 1. Kim[16] et al. | air/BP/300nm SiO₂/Si | (A) 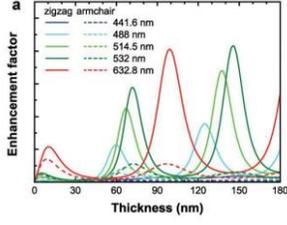 | (B) 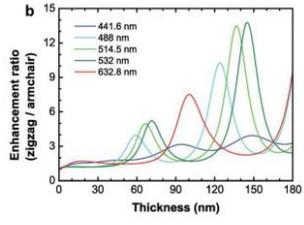 |
| 2. Our calculations | air/BP/300nm SiO₂/Si | (C) 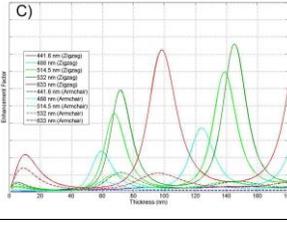 | (D) 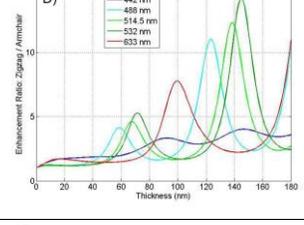 |
| 3. Our work | air/BP/0.1mm polyimide | (E) 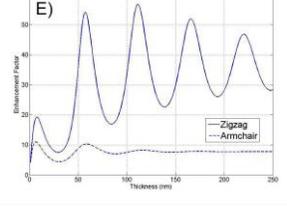 | (F) 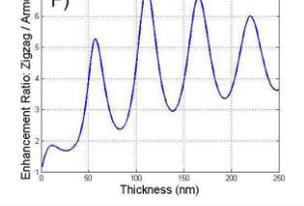 |

**Table S11** Comparison between the reported calculation of interference factors and our calculation results for different laser wavelengths: (A) and (B) are interference factors reported by Kim[16] et al, (C) - (F) are our calculation results.

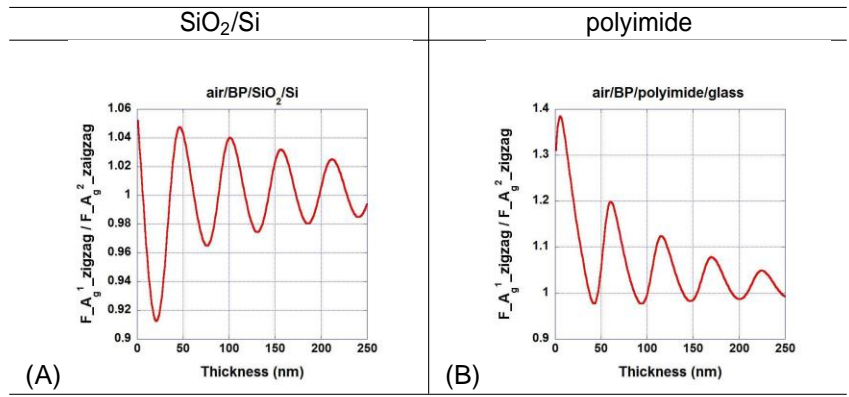

| SiO₂/Si | polyimide |
|---|---|
| (A) | (B) |

**Table S12** Fig.(A): The ratio of enhancement factor of $A^1_g$ mode over that of $A^2_g$ mode when excitation was parallel to zigzag direction of BP, with BP on SiO₂/Si substrate. (B): with BP on polyimide substrate.

**6. Calculation of amplitudes of Raman tensors |a| and |c|**

Basing on the multi-reflection model (MRM) (developed from Fresnel's law and extensively applied in the studies of 2D materials)[16,48,47], measured Raman intensity can be written as:

$$R = I \times F; \tag{4}$$

where the $I$ is the intrinsic Raman intensity considering only the electron−phonon and phonon−electron interaction, while $F$ is the total interference enhancement factor. The enhancement factor of incidence when it's parallel to zigzag direction is larger than (no more than 3.5 times for SiO$_2$/Si substrate and 5.5 times for polyimide substrate with BP's thickness from 0 to 100 nm) that when it's parallel to armchair direction, suggesting the interference effect to be stronger in zz (zigzag) configuration, but it's still small in comparison with that of longer wavelength. Our results are also in agreement with previous work,[16] the ratio of enhancement factors $\frac{F_{zigzag}}{F_{armchair}}$ didn't change rapidly as growing thickness as well as showed far smaller values.

$I$ was calculated after dividing $R$ by $F$. Also, $I$ can be described by this equation: $\mathrm{I} = |e_i \cdot \mathbb{R} \cdot e_s|^2$, where $\mathbb{R}$ is the Raman tensor, $e_i$ and $e_s$ are the light polarization vectors of the incident and scattering beams respectively. From group theory, the Raman tensor for A$_g$ modes of BP in the backscattering geometry is:

$$\mathbb{R}(A_g) = \begin{pmatrix} a & 0 & 0 \\ 0 & b & 0 \\ 0 & 0 & c \end{pmatrix} \tag{5}$$

in which

$$a = |a|e^{i\varphi_a}; c = |c|e^{i\varphi_c}; \tag{6}$$

$$\varphi_a = arctg[\frac{\frac{\partial \varepsilon'_{xx}}{\partial q^{A_g}}}{\frac{\partial \varepsilon_{xx}}{\partial q^{A_g}}}], \varphi_c = arctg[\frac{\frac{\partial \varepsilon'_{zz}}{\partial q^{A_g}}}{\frac{\partial \varepsilon_{zz}}{\partial q^{A_g}}}], \tag{7}$$

$$\varepsilon = \varepsilon^f + i\varepsilon"; \tag{8}$$

Here |a| and |c| are amplitude factors of Raman tensors A and C; $\varphi_a$ and $\varphi_c$ are the phases of Raman tensor elements; $q^{A_g}$ are the normal coordinates of the Raman modes, and $\varepsilon^f$ is the imaginary part of complex relative dielectric complex $\varepsilon$, relating to the absorption of BP, which can be expressed as a function of excitation energy:[29]

After that, we follow the transformation of the Raman tensors from crystalline orientation of BP to the rotated coordinate frame[18] so that the Raman tensors can be expressed as:

$$\mathbb{R}(A_g) = \begin{pmatrix} a\sin^2\theta & 0 & \frac{1}{2}(a-c)\sin2\theta \\ 0 & b & 0 \\ \frac{1}{2}(a-c)\sin2\theta & 0 & a\cos^2\theta + c\sin^2\theta \end{pmatrix} \tag{9}$$

Here to mention that, there are two notations for Raman scattering efficiency of A$_g$ modes under parallel configurations:[16,18]

$$I_{A_g} = (a \cdot \sin^2\theta + c \cdot \cos^2\theta)^2; \tag{10}$$

or[17,19]

$$I_{A_g} = (a \cdot \cos^2\theta^f + c \cdot \sin^2\theta)^2; \tag{11}$$

Eq.(8) and (9) are both correct. The difference is introduced by different definitions of $e_i$ and $e_s$:

in Eq.(8):

$$e_i = e_s = (\sin\theta, 0, \cos\theta);$$

while $\theta$ here is defined as the angle between x (zigzag) and $e_s$;

in Eq.(9):

$$e_i = e_s = (\cos\theta^f, 0, \sin\theta^f);$$

while $\theta^f$ is the angle between z (armchair) and $e_s$;

Hence that the forms of the amplitude ratio: $|\frac{c}{a}|$ calculated from Eq.(8) and $|\frac{a}{c}|$ from Eq.(9) are different but the values of them are the same. In this work, under parallel configuration, the polarization vectors are given by $e_i = e_s = (\sin\theta, 0, \cos\theta)$, where the angle $\theta$ is measured with respect to the zigzag direction. Then the Raman intensity can be written as[17,19]:

$$I_{A_g} = |(\sin\theta, 0, \cos\theta) \cdot \overline{R} \cdot (\sin\theta, 0, \cos\theta)^T|^2 = |a|^2[(\sin^2\theta + |\frac{c}{a}| \cdot \cos\Phi_{ca}\cos^2\theta)^2 + |\frac{c}{a}|^2 \cdot \sin^2\Phi_{ca}\cos^4\theta]; \tag{12}$$

where the term $\Phi_{ca}$ is the phase difference of $\varphi_c$ - $\varphi_a$, $\theta$ denotes the angle between the polarization of incident light and zigzag direction. The amplitude ratio $|\frac{c}{a}|$ and phase difference $\Phi_{ca}$ are two parameters of anisotropy in BP.

| Wavelength (nm) | Laser energy (eV) | $A_g^1$ | | | | | | | | |
| | | Thickness (nm) | | | $\left|\frac{c}{a}\right|$ | | | $\Phi_{ca}(°)$ | | |
| | | Ref.[19] | Ref.[16] | *Ref.[17] | Ref.[19] | Ref.[16] | *Ref.[17] | Ref.[19] | Ref.[16] | Ref.[17] |
| 785 | | | | 5 - 200 | | | | | | |
| 633 | 1.95 | 360 | 90 | 5 - 200 | 2.00 | 0.98 | 0.6 - 1.3 | 64 | 47 | 0 - 70 |
| 532 | 2.33 | 360 | 90 | 5 - 200 | 1.96 | 0.75 | | 0 | 43 | |
| 514.5 | 2.41 | | 5, 65, 75, 90 | | | 0.76 | | 0 | 34 | |
| 488 | 2.54 | 360 | 90 | | 3.43 | 1.20 | | | 48 | |
| 442 | 2.80 | | 5, 65, 75, 90 | | | 4.01 | | | 0 | |

| Wavelength (nm) | Laser energy (eV) | $A_g^2$ | | | | | | | | |
| | | Thickness (nm) | | | $\left|\frac{c}{a}\right|$ | | | $\Phi_{ca}(°)$ | | |
| | | Ref.[19] | Ref.[16] | *Ref.[17] | Ref.[19] | Ref.[16] | *Ref.[17] | Ref.[19] | Ref.[16] | Ref.[17] |
| 785 | | | | 5 - 200 | | | | | | |
| 633 | 1.95 | 360 | 90 | 5 - 200 | 1.32 | 0.79 | 0.3 - 0.95 | 98 | 91 | 50 - 90 |
| 532 | 2.33 | 360 | 90 | 5 - 200 | 1.33 | 0.52 | | 95 | 120 | |
| 514.5 | 2.41 | | 5, 65, 75, 90 | | | 0.50 | | 90 | 111 | |
| 488 | 2.54 | 360 | 90 | | 0.76 | 0.47 | | | 97 | |
| 442 | 2.80 | | 5, 65, 75, 90 | | | 0.24 | | | 179 | |

**Table S13** A summary in previous studies[16,17,19] on laser energy dependence of the absolute values of Raman tensor ratio $\left|\frac{c}{a}\right|$ and phase difference $\Phi_{ca}$ of $A_g^1$ and $A_g^2$ modes; in addition, the values and range of thickness of BP flakes used for those studies were also collected.

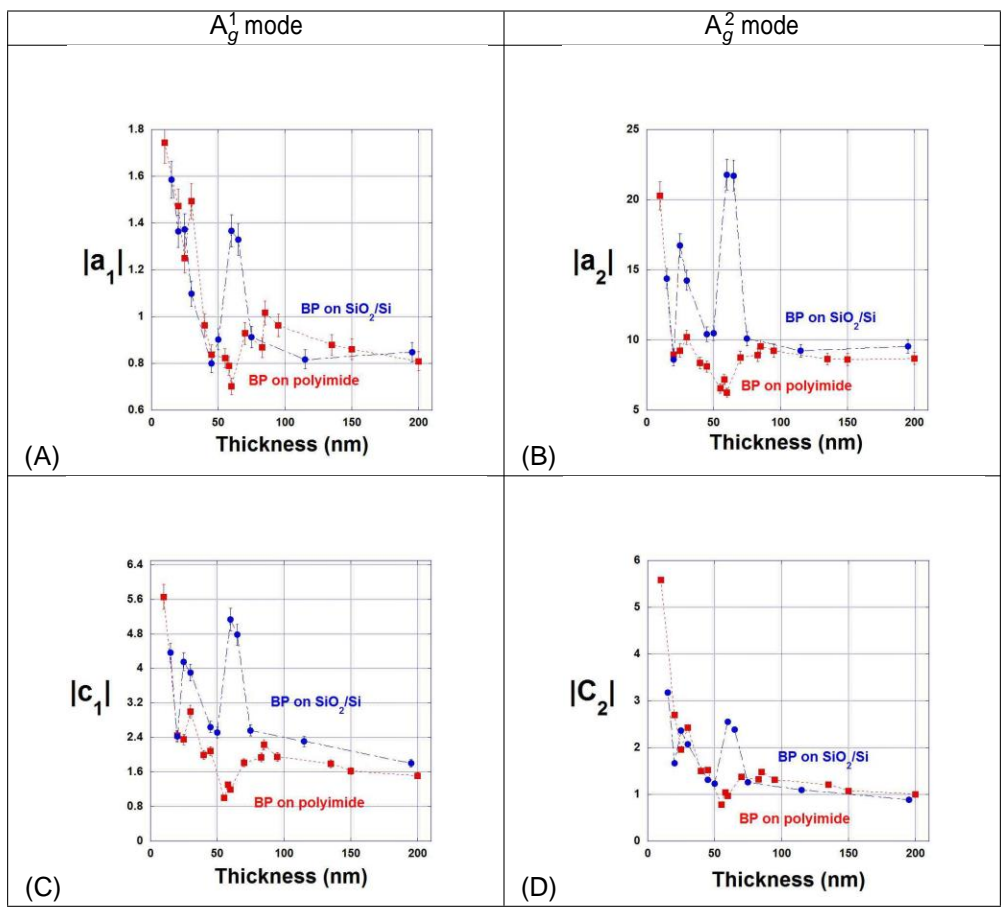

**Table S14** Fig.(A)−(D): the ratio of amplitudes of Raman tensors c and a of $A_g^1$ and $A_g^2$ mode as functions of thickness; the red solid cubes are data of BP samples on polyimide substrate, while those of blue solid dots are data of BP samples on $SiO_2$/Si substrate.

**7. $A_g^1$ mode was too weak to be observable when excitation polarization was parallel to zigzag direction of BP.**

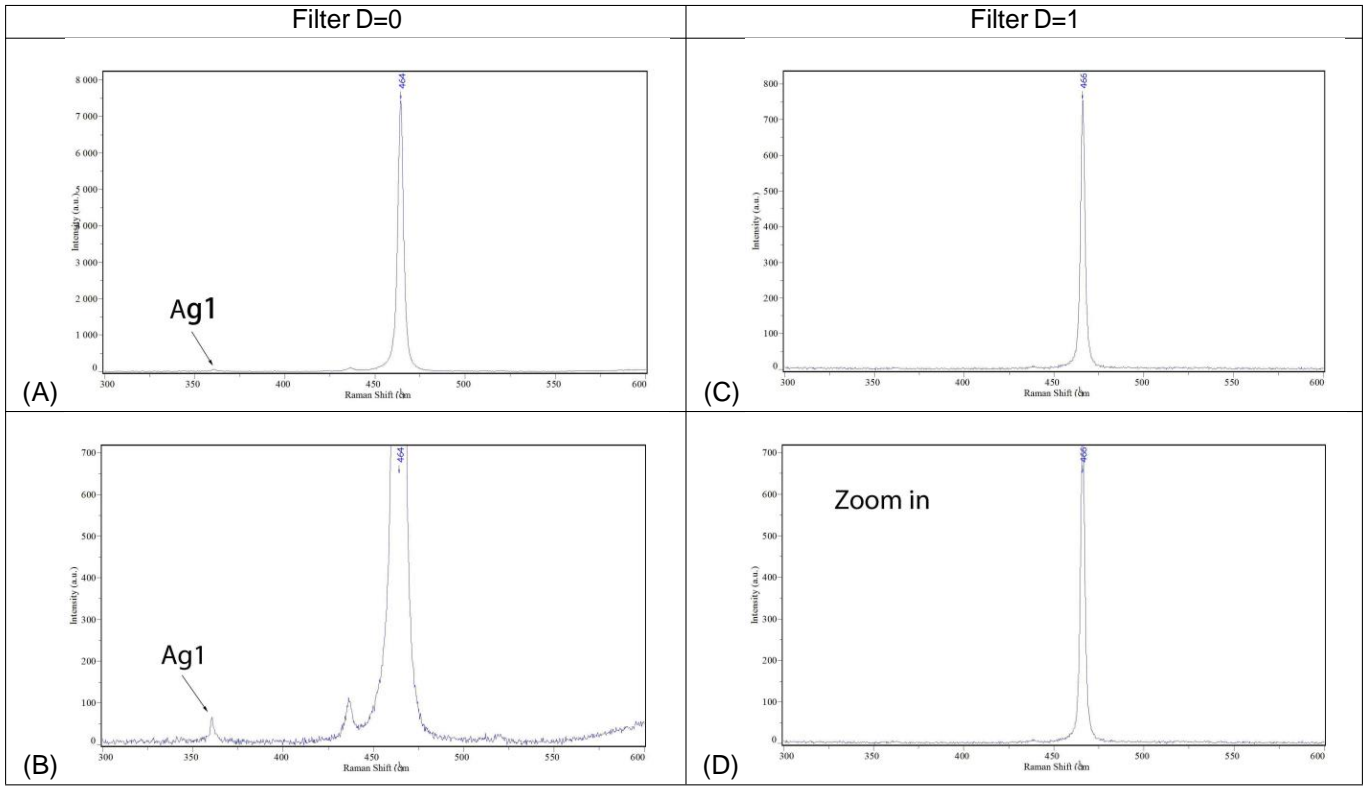

**Table S15** (A) and (B) are Raman spectra of 50-nm-thick BP sample on SiO$_2$/Si, with filter D=0, A; (C) and (B) are Raman spectra measured using filter D=1;

Here, we compared the Raman profiles of 50-nm-thick BP sample on SiO$_2$/Si substrate measured when incidence was along armchair direction using filter D=1 and D=0. $A_g^1$ mode can be detected if using no filter (D=0) while that failed if using filter D=1, in that when the incidence is along armchair direction, $A_g^1$ will be too week to be visible because of the interference of the intense Raman intensity of $A_g^2$ mode. Therefore, in consideration of avoiding laser heating effect, we chose filter D=1 to reduce the laser power for ARPRS measurements in our study as well as treated the peak intensity of $A_g^1$ mode to be that of substrate for calculations when the excitation polarization was parallel to armchair direction of BP.